\documentclass[a4paper,reprint,prd,aps,floats,floatfix,showpacs,twocolumn,superscriptaddress,nofootinbib,nolongbibliography, noeprint]{revtex4-2}
\usepackage{textcomp}
\usepackage{amssymb,amsmath,amsfonts,bigints}
\usepackage{microtype,gensymb,mathtools}
\usepackage{verbatim,needspace,enumitem,etoolbox,afterpage,tabularx,xspace,siunitx}
\usepackage{aasmacros}
\usepackage{graphicx}
\usepackage{dcolumn}
\usepackage{bm}
\usepackage{lipsum}
\usepackage{physics}
\usepackage[normalem]{ulem}
\usepackage[dvipsnames, table]{xcolor}
\usepackage{tikz}
\usetikzlibrary{
    arrows.meta,
    positioning,
    shapes.geometric,
    fit,
    backgrounds
}

\definecolor{linkcolor}{rgb}{0.0,0.3,0.5}
\usepackage[colorlinks = true,
            linkcolor = linkcolor,
            urlcolor  = linkcolor,
            citecolor = linkcolor,
            anchorcolor = linkcolor]{hyperref}
\usepackage[capitalize]{cleveref}
\Crefname{equation}{Eq.}{Eqs.}
\Crefname{figure}{Fig.}{Figs.}
\Crefname{tabular}{Tab.}{Tabs.}
\Crefname{section}{Sec.}{Secs.}
\Crefname{subsection}{Sec.}{Secs.}
\Crefname{appendix}{App.}{Apps.}

\usepackage{array}

\newcommand{\msun}{\mathrm{M_{\odot}}}

\newcommand{\mc}{\mathcal{M}_{\mathrm{c}}}
\newcommand{\tc}{t_{\mathrm{c}}}
\newcommand{\psd}{\langle \widehat{\mathrm{PSD}} \rangle_\mathfrak{s}}
\newcommand{\fap}{\mathrm{FAP}}
\newcommand{\faptr}{\mathrm{FAP}_{\mathrm{thr}}}
\newcommand{\Ns}{N_{\mathfrak{s}}}

\usepackage{orcidlink}
\begin{document}
\sisetup{range-phrase=-, range-units=single}

\newcommand{\insubria}{\affiliation{Como Lake Centre for Astrophysics, DiSAT, Università dell’Insubria, via Valleggio 11, I-22100 Como, Italy}}
\newcommand{\milan}{\affiliation{Dipartimento di Fisica “G. Occhialini”, Università degli Studi di Milano-Bicocca, Piazza della Scienza 3, 20126 Milano, Italy}}
\newcommand{\infn}{\affiliation{INFN, Sezione di Milano-Bicocca, Piazza della Scienza 3, 20126 Milano, Italy}}
\newcommand{\bham}{\affiliation{Institute for Gravitational Wave Astronomy \& School of Physics and Astronomy, University of Birmingham, Birmingham, B15 2TT, UK}}
\newcommand{\oca}{\affiliation{Université Côte d’Azur, Observatoire de la Côte d’Azur,
CNRS Artemis, Bd de l’Observatoire, F-06304 Nice, France}}

\author{Francesco~Nobili*~\orcidlink{0000-0002-7387-6754}}
\insubria
\infn
\email{fnobili@uninsubria.it}
\date{\today}

\author{Malvina~Bellotti~\orcidlink{0000-0002-7387-6754}}
\milan
\oca

\author{\\Riccardo~Buscicchio~\orcidlink{0000-0002-7387-6754}}
\milan
\infn
\bham

\author{Massimo~Dotti~\orcidlink{0000-0002-7387-6754}}
\milan 
\infn

\author{Alessandro~Lupi~\orcidlink{0000-0002-7387-6754}}
\insubria
\infn

\title{Fast premerger detection of massive black-hole binaries in LISA based on time-frequency excess power}

\begin{abstract}

The Laser Interferometer Space Antenna is expected to observe gravitational waves from massive black hole binaries across cosmic time. Many are anticipated to be detectable hours to weeks before coalescence. 
We present a fast algorithm for the premerger detection and preliminary characterization of such binaries. 
The method performs a search for excess power with a chirping time-frequency morphology in short-time Fourier transform spectrograms. 
By tiling the time-frequency plane with slices defined by the quadrupole frequency evolution, we define a signal significance relative to a fitted background distribution of instrumental noise and Galactic foreground. 
Individual search triggers are followed by a coherence tracker that groups triggers consistent with the same physical signal over time. Doing so, our analysis provides progressively refined estimates of the chirp mass and coalescence time. 
We validate our algorithm on the SangriaHM LISA Data Challenge dataset, successfully detecting all 15 injected massive black-hole binaries: 14 of them hours to weeks before merger, while one is only detected after the binary coalescence.
The algorithm yields chirp mass relative errors below $3\%$ for high-SNR sources and coalescence time uncertainties of up to a few hours.
With a computational cost of less than a second to process a 10-day data segment on a single core, our approach is suitable for generating real-time alerts, triggering protected observational periods, and providing informative priors for Bayesian parameter estimation.
\end{abstract}

\maketitle

\section{\label{sec:intro} Introduction}

The Laser Interferometer Space Antenna (LISA) will be the first space-based gravitational-wave observatory (GW)~\cite{2017arXiv170200786A}. 
Operating in the millihertz regime, LISA will bridge the observational gap between current and future ground-based interferometers~\cite{2020LRR....23....3A, 2025arXiv250312263A, 2019BAAS...51g..35R} and pulsar timing array observatories~\cite{2016MNRAS.458.1267V}. 
A wide variety of sources is expected to be observed~\cite{2024arXiv240207571C,2023LRR....26....2A}: the multitude of Galactic double white dwarfs (DWDs) producing a stochastic foreground, stellar-mass black holes in their early inspiral, compact stellar remnants orbiting massive black holes (i.e. extreme mass-ratio inspirals), and coalescing massive black-hole binaries (MBHBs).
The last are characterized by loud transient signals lasting from hours to weeks in the LISA sensitivity band.

The simultaneous presence of these sources, overlapping in time and frequency, makes the LISA data analysis a complex challenge motivating the development of global fits~\cite{Deng:2025wgk, Katz:2024oqg, Littenberg:2023xpl, Strub:2024kbe, Astorino:2025ccl}: pipelines capable of performing simultaneous Bayesian parameter estimation on an uncertain number of sources within the data stream. 
Owing to their complexity, such algorithms likely require months of accumulated data before delivering accurate results.

Nonetheless, a fast detection and preliminary characterization of transient sources is needed for time-critical scientific investigations.
This is the objective of low-latency pipelines, designed to operate in real time and release rapid alerts containing preliminary source-property estimates within one hour from the reception of new data on the ground.
The target alert emission timescale is valuable to trigger protected observational periods preventing scheduled gaps in the data stream, and to enable multimessenger observations with electromagnetic observatories. Similar low-latency strategies have been extensively developed for ground-based detectors~\cite{Singer:2015ema, Marx:2024wjt}, primarily targeting neutron star mergers, and have already enabled groundbreaking results~\cite{LIGOScientific:2017adf, LIGOScientific:2017vwq, LIGOScientific:2017ync, Hallinan:2017woc, Troja:2017nqp}.
Among LISA detectable sources, MBHBs are the most promising candidates for such followups. 
At present, massive black holes observable by LISA (i.e. in the $10^4$ -- $10^7 \msun$ mass range) are identified through electromagnetic observations, either as the bright central engines of Seyfert galaxy nuclei \citep[see][for a review]{1977ARA&A..15...69W} or as responsible for the dynamics of nuclear stars \citep[as for the case of Sgr A*, e.g.][]{GRAVITY:2018ofz} and of megamaser discs \citep{2016ApJ...819...11V}.
Mergers of MBHBs occurring in gas-rich environments may produce peculiar electromagnetic signatures both before and after coalescence~\cite{2022LRR....25....3B}. 
While no clear electromagnetic evidence for MBHBs have been presented so far\footnote{The best candidate so far is represented by the pair of radio cores at $\approx 7$ pc projected separation discussed in~\cite{2006ApJ...646...49R}.}, several strategies have been proposed to perform multimessenger astronomy through LISA GW detections and several EM observatories~\cite{Mangiagli:2022niy, Saini:2022hrs, Lops:2022ooj, DottiESA25}. 
Notably, searches for MBHB signatures in LSST photometric data~\cite{LSST:2008ijt, Xin:2021mmk, Xin:2024fci, Xin:2025voy}, complemented with spectroscopic validations~\cite{Bertassi:2025vry}, and follow-up observations with the NewAthena telescope~\cite{2023MNRAS.521.2577P, Piro:2021oaa} have been put forward in the literature.

Several methods have been proposed for the rapid detection and characterization of GWs from MBHBs.
One class of such methods aims at accelerating the analysis of full inspiral-merger-ringdown MBHB signals, mainly relying on approximate detection and parameter estimation schemes~\cite{Cornish:2021smq, Katz:2021uax, Weaving:2023fji, Hoy:2024ovd, Sharma:2024sfb, Jan:2024zhr, ElGammal:2025dkz, Deng:2025qhx}. 
A second class focuses on the MBHB premerger signals, leveraging either approximated matched-filtering schemes~\cite{CabournDavies:2024hea, Deng:2025qhx,Tenorio:2025gci} or machine-learning approaches~\cite{Ruan:2024qch, Houba:2024mqj, Isfan:2025nhb}. In the context of LISA, several time–frequency frameworks have recently been proposed for the search and characterization of massive black hole binaries~\cite{Tenorio:2025gci}, extreme mass-ratio inspirals~\cite{Speri:2025ucn}, and stellar-mass black holes~\cite{Bandopadhyay:2025fyx}. Furthermore, novel time-delay interferometry techniques are currently being developed with the objective of pinpointing the sky position of MBHBs~\cite{Barroso:2024azl}. Finally, several time-frequency (TF) excess power methods have been presented in past literature, especially regarding agnostic burst searches~\cite{Chatterji:2004qg, Klimenko:2015ypf, Martini:2025ewk}, or waveform-free estimation of eccentricity~\cite{Tai:2014bfa, Bose:2021pcw, Fernandes:2026dqy} applied to data from ground-based interferometers.

In this work, we present a fast detection algorithm based on the search for excess power with a time-frequency {}{} chirping morphology through short-time Fourier transform (STFT) spectrograms.
We do so inspired by the successful application of similar methods to ground-based interferometers data, where analogous detection techniques using the same data representation have been proposed~\cite{Tenorio:2021wmz, 2004CQGra..21S1809C, 2008CQGra..25k4029K, 2012JPhCS.363a2032N, Cornish:2014kda, klimenkoCWBPipelineLibrary2021, 2023PhRvD.107f2002S, Licciardi:2024bhv}.
However, we highlight two key differences with respect to our method in the LISA context:
(i) our algorithm is able to identify multiple overlapping signals in a single chunk of observation, and (ii) it tracks parameter estimates and uncertainties, progressively updating them as further data are acquired.

Datastreams of current ground-based detectors are largely noise-dominated. By contrast, LISA data are signal-dominated, with a background composed of stationary instrumental noise and a cyclostationary foreground from Galactic DWDs~\cite{2009CQGra..26i4030N, Buscicchio:2025zeb, Pozzoli:2024wfe}. 
We explicitly account for this through off-source noise estimation.

Overlapping signals in current ground-based detectors are extremely rare. On the contrary, long-duration LISA MBHB inspirals may overlap in time~\cite{2021MNRAS.507.5069A}.
As we shall see in following sections, we make a considerable effort to retain large flexibility in the selection of spectrogram pixels used for statistical testing. The motivation is two-fold:
the extended duration of MBHB signals makes them susceptible to contamination from glitches and gaps. 
Glitches are loud and brief transients that have been observed in the data of the LISA Pathfinder mission~\cite{Armano:2016bkm, Armano:2018kix}. 
Their characterization is currently a live research topic~\cite{2025CQGra..42f5018C, Spadaro:2023muy, 2022PhRvD.106f2001A}. 
Gaps are downtime periods during which LISA will not collect science-quality data~\cite{Burke:2025bun}, either due to scheduled maintenance periods, or to unscheduled -- temporary or permanent -- data losses. 
Working in the TF domain is particularly advantageous to handle both artifacts, given the large flexibility in selecting spectrogram pixels unaffected by them.

The paper is organized as follows.
In~\cref{sec:methodology}, we describe our algorithm: we detail our time–frequency data representation and noise estimation, both based on STFTs, in~\cref{subsec:tf_representation}; 
in~\cref{subsec:detection_strategy}, we introduce the signal model assumed and the procedure to estimate the detector-frame chirp mass $\mc$ and coalescence time $\tc$ from each excess-power detection; 
in~\cref{subsec:coherent_refinement} we describe how search triggers from each segment are coherently tracked through time to identify physical sources and iteratively update their chirp mass and time-to-coalescence estimates.
In~\cref{sec:results}, we summarize our results, showing that our algorithm successfully detects all sources in the SangriaHM dataset with minimal computational cost.
Finally, we conclude in~\cref{sec:conlusions} with a discussion of the limitations of our approach and possible extensions for future work.
In~\cref{app:estimates_all_sangria} and~\cref{app:background_distro_derivation} we illustrate results for each individual system, and report a detailed derivation of the background statistics, respectively.

\section{\label{sec:methodology} Methodology}

\subsection{\label{subsec:tf_representation} Time-frequency representation}

\begin{figure}[t]
    \includegraphics[width=\columnwidth]{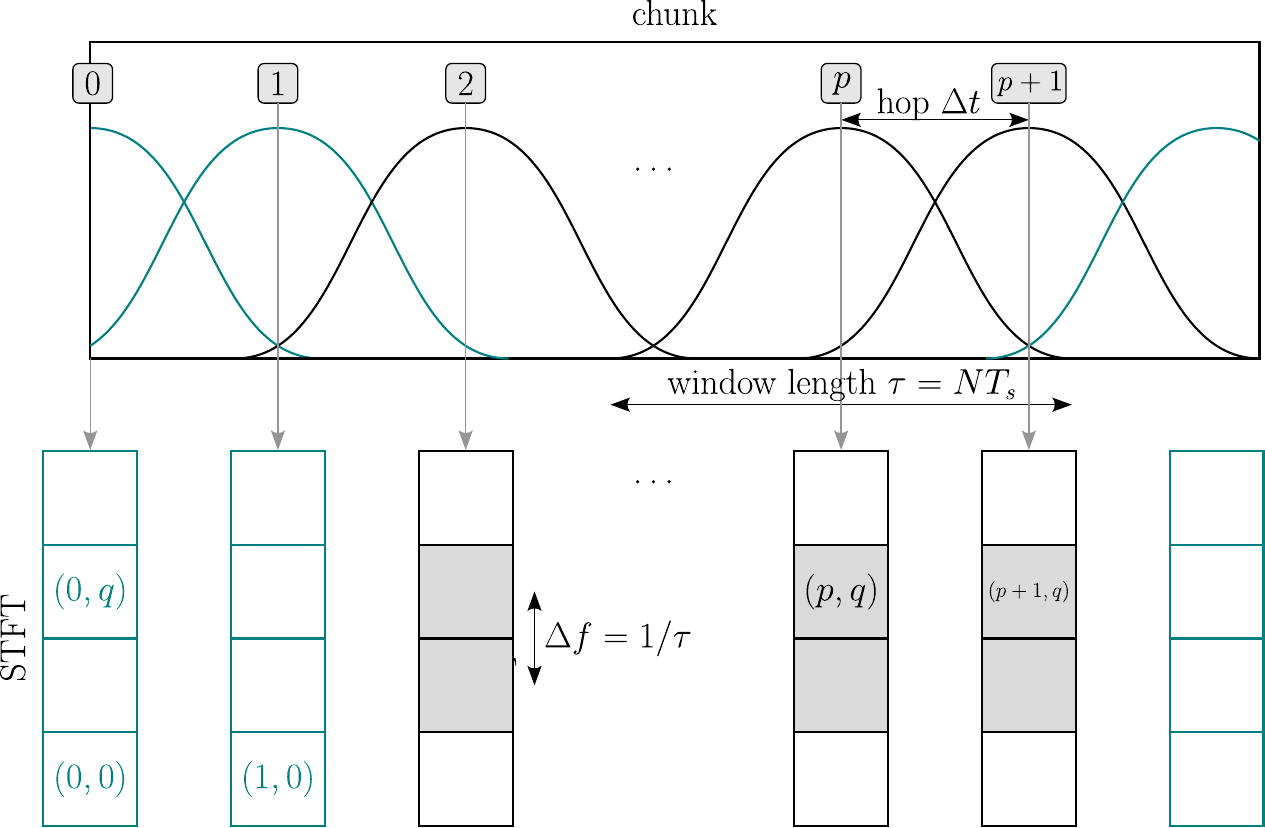}
    \caption{Summary of STFT scheme and free parameters. In our default configuration we consider a sliding chunk of 10 days sampled with cadence $T_s=5$ s. Each chunk is uniquely associated to a spectrogram, as defined in Eqs.~\eqref{eq:discreteSTFT} and~\eqref{eq:stftpower}: segments of length $\tau$, overlapping by $95\%$ are Hann-windowed according to Eq.~\eqref{eq:hannwindow} and Fourier transformed, yielding a frequency array associated to the time $p\Delta t$. Note that windows in the figure are overlapping by $50\%$ for a cleaner visualization. We further {} mask out pixels corresponding to windows with support outside of the chunk borders (teal-bordered boxes), and pixels above $10^{-3} {\rm Hz}$ and below $10^{-4} {\rm Hz}$ (white, black-bordered boxes).}
    \label{fig:Figure1}
\end{figure}

Throughout our work, we assume LISA inter-spacecraft laser-phase measurements to be suitably delayed and combined into time-delay interferometric variables (TDI), to achieve the target GW sensitivity through synthetic laser-noise cancellation~\cite{Tinto:2020fcc}.
We do so coherently using the \textit{SangriaHM} LISA Data Challenge~\cite{lejeuneLISADataChallenge2022,2022arXiv220412142B}. 
The simplest TDI combinations, TDI-1.0, assume a stationary array described by three constant armlengths, so that the light travel time along a given arm is the same in both propagation directions. The dataset is instead released in the form of TDI-1.5 observables, which account for the Sagnac effect induced by the rotation of the constellation — implying that the up and down optical paths are unequal — thus involving six distinct, though still time-independent, delays. Future simulated datasets will drop this approximation in favor of TDI-2.0 combinations, which further account for time-dependent armlengths~\cite{Tinto:2020fcc, Tinto:2022zmf}.
As we shall see in~\cref{subsec:tf_representation}, this would only affect the whitening procedure.
The dataset is released in the form of $X$, $Y$, and $Z$ variables. 
Through linear combinations of such TDI variables, we construct auxiliary, uncorrelated ones (known in literature as $A$, $E$, and $T$).
The dataset features one year of observations with a cadence of $T_s = 5 {\rm s}$, containing simulated instrumental noise~\cite{LISA2018}, bright and foreground signals from a simulated Galaxy of white dwarf binaries, and 15 MBHB merger signals following the \textsc{PhenomHM} waveform model~\cite{Khan:2015jqa}. 
Henceforth, similarly to the convention adopted in Ref.~\cite{Deng:2025qhx}, we will refer to the $n$-th MBHB system, when ordered by increasing merger time, as MBHB-$n$ (starting with $n=0$).

Our algorithm is designed to operate in real time, processing new data as soon as they become available. 
To avoid ambiguity, we refer to the interval of time-domain data considered at each step as \textit{chunk}. 
The choice of chunk length is adjustable depending on the typical source lifespan within the LISA band. 
We adopt a default time-sliding chunk of fixed $10$ days length, so that incoming data points replace older ones, sliding forward every 200 minutes.

Once a data chunk is collected, we compute its TF representation via STFT, as implemented in \textsc{Scipy}~\cite{scipySTFT}. 
For a discrete time series $x[k]$ sampled with cadence $T_s$, the STFT is obtained by taking the discrete Fourier transform (DFT) of short, overlapping segments. 
Each segment is selected multiplying the data by a window function $w[k - ph]$ centered at time $p \Delta t$, and of duration $\tau = N T_s$, with $p$ denoting the segment number, and $N$ the number of samples in each segment.
The window is then shifted forward by $h$ samples at each step, corresponding to a time-shift (or \emph{hop}) of $\Delta t = h T_s$.
The STFT spectrum evaluated at time $p \Delta t$ and frequency $q \Delta f$ is therefore
\begin{equation}
X[p,q] = \sum_{k=0}^{N-1} x[k]\, w[k - ph]\, e^{-i 2 \pi q k / N}. \label{eq:discreteSTFT}
\end{equation}
In this work, we use the term \textit{spectrogram} to denote the discrete-frequency power spectral density (PSD), defined as
\begin{equation}
\mathrm{PSD}[p,q] = T_s \frac{\left|X[p,q] \right|^2}{\left\lVert w\right\lVert^2} , \label{eq:stftpower}
\end{equation}
which has units of $\mathrm{Hz}^{-1}$. Here, $\left\lVert w\right\lVert$ is a normalization constant defined as
\begin{equation}
\left\lVert w\right\lVert^2 = \sum_{k=0}^{N-1} \left| w[k] \right|^2 . \label{eq:stftnormalization}
\end{equation}
We employ the standard Hann window, i.e., a scaled cosine with support over $0 \leq n < N$:
\begin{equation}
w[n] = \frac{1}{2} - \frac{1}{2} \cos\left( \frac{2\pi n}{N-1} \right) . \label{eq:hannwindow}
\end{equation}
This window is widely used as it provides a good compromise between frequency resolution and dynamic range~\cite{1981ITASS..29...84N}.

\begin{figure}[t]
    \centering
    \includegraphics[width=1\columnwidth]{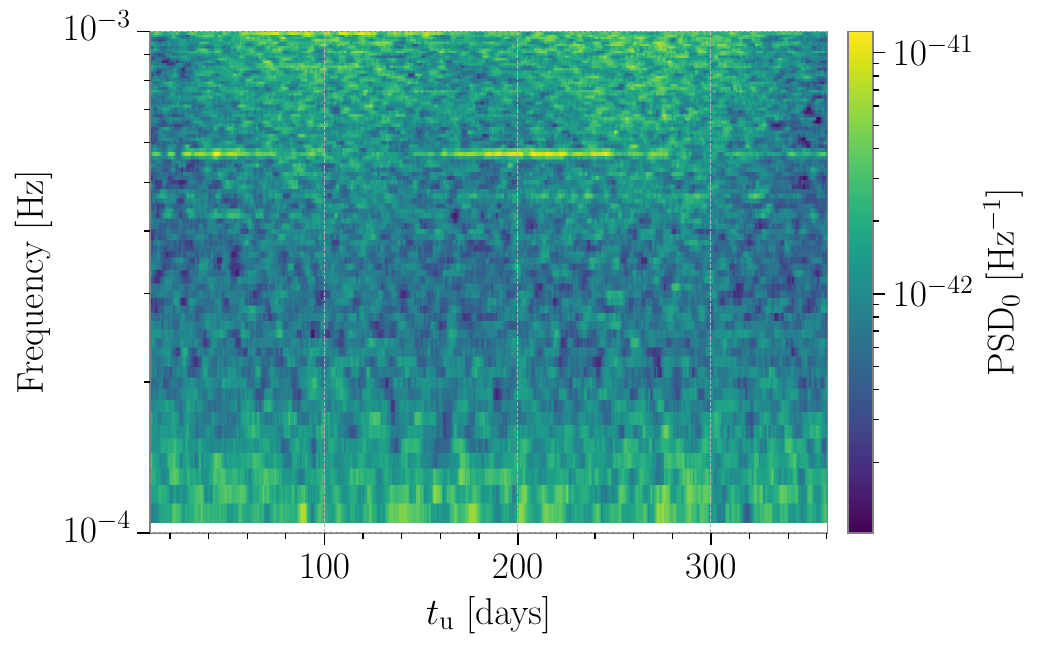}
    \caption{Array of PSDs used for whitening the spectrograms at each update time $t_{\rm u}$. The Galactic confusion noise modulation is visible as a periodic brightening particularly clear above $0.5 {\rm mHz}$. 
    The brightness-modulated narrow horizontal line at $f=\num{5.6e-4}$ Hz is the signal emitted by a loud DWD source, at a distance of $0.1 {\rm kpc}$, far from the Galactic center, which is successfully suppressed in the whitened spectrograms.}
    \label{fig:Figure2}
\end{figure}

In our default configuration, we employ segments of duration $\tau=10^5$ s with a $95\%$ overlap, restricting the analysis to the frequency range between $10^{-4}$ and $10^{-3}$ Hz. 
While extending the search to lower frequency bands could enhance the early detection of lower-mass signals characterized by prominent inspiral phases~\cite{CabournDavies:2024hea}, we adopt this conservative lower limit to demonstrate that our algorithm remains effective even under restrictive conditions.

We show in~\cref{fig:Figure1} the overall STFT scheme mapping each windowed segment to a Fourier series:
to construct a spectrogram, we consider only segments whose window do not overlap with chunk borders. 
For a $10$ days-long chunk, this amounts to consider segments centered from $0.58$ up to $9.38$ days.
Doing so, we drop zero-padded segments as constructed in the \texttt{scipy.signal.ShortTimeFFT} implementation~\cite{scipySlidingWindows}, thus
avoiding ringing artifacts due to the Gibbs phenomenon~\cite{gottliebReviewDavidGottliebs2011} near sharp discontinuities, which in this case happen at data chunk borders.

We further process each spectrogram through whitening.
In ground-based detectors literature, methods for both on-source and off-source noise estimation have been devised~\cite{Chatziioannou:2019zvs, Isi:2020uxj}.
In LISA transient searches, a major component of the background arises from Galactic sources, whose detector modulation becomes significant over weeks timescales~\cite{Buscicchio:2025zeb}.
Therefore, background PSD can be calibrated within reasonable approximation using data extending over the same time in the past, in absence of other bright transient signals.

Low-latency pipelines running on the LISA Distributed Data Processing center (DDPC) will likely be equipped with regular updates of instrumental noise, DWD confusion noise, and resolved DWDs parameter estimates, which together define the noise-hypothesis distribution, ${\cal H}_0$. 
These updates may be obtained by dedicated blocks of the ground-segment global-fit pipeline, sharing background point estimates previous to their full convergence. Such preliminary estimates of the $PSD$ of the background process may be directly used in the whitening procedure described below, and in the computation of the distribution of the detection statistic introduced in Sec.~\ref{subsec:detection_strategy}, under the background assumption ${\cal H}_0$.  However, an independent estimate of the background performed in the context of low-latency searches may prove beneficial, adding redundancy in the analysis chain. 
While this objective may be investigated in future work, in our analysis we estimate the background PSD directly from the SangriaHM dataset containing only instrumental noise and Galactic sources.

We consider 500 reference update times $t_u^{(k)}$ across the year of simulated data, i.e.
\begin{align}
t^{(k)}_{\rm u} &= 10 {\rm d} + k \times 0.71\,{\rm d}
\end{align}
At each update, we construct a reference PSD estimator from a 5 days-long chunk of background data $X_{\rm bkg}$ in $\left[t_\mathrm{u} -5\mathrm{d}, t_\mathrm{u}\right]$, mimicking an estimate based on previous data, as provided by the dedicated global fit blocks.
We do so by computing STFTs as in~\cref{eq:stftpower}, with windowing and overlap parameters as for the data, and taking the median value over the 76 time segments available to each $t_{\rm u}$
\begin{equation}
\mathrm{PSD}_0[q; t_{\rm u}] = \frac{T_s}{{\left\lVert w\right\lVert^2}}\operatorname*{median}_{p\Delta t \in [t_{\rm u}-5{\rm d}, t_{\rm u}]} {\left|X_{\rm bkg}[p,q] \right|^2}.\label{eq:psd0qtu}
\end{equation}

We therefore obtain 500 PSD estimates ${\rm PSD}_0[q;t_\mathrm{u}]$, each one a vector labelled by a reference $t_u$, whose frequency-content evolution over the year is shown in ~\cref{fig:Figure2}.
Along with the Galactic confusion noise, this procedure naturally incorporates into the background model pixel power from faint or not-yet-resolved signals. 
In~\cref{fig:Figure2} such a scenario is represented by a bright DWD source at $f=0.56~{\rm mHz}$. 
\begin{figure}[t]
    \includegraphics[width=\columnwidth]{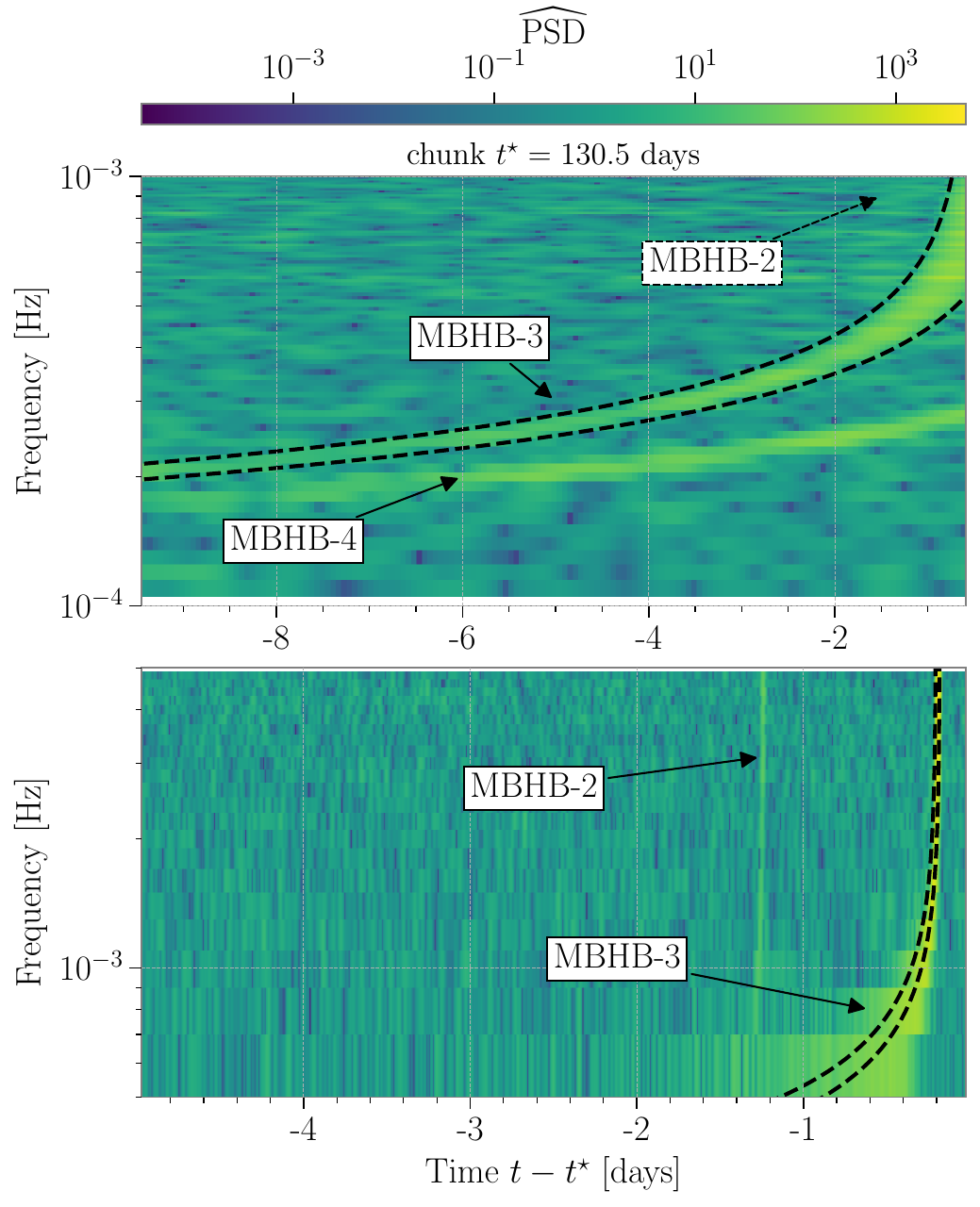}
    \caption{
    The top panel shows the default setup described in~\cref{subsec:tf_representation}, while the bottom panel illustrates the configuration optimized for the detection of MBHB-2, discussed in~\cref{sec:results}. This source has a chirp mass of $1.2 \times 10^6 M_\odot$: it is therefore characterized by a fainter inspiral and a big portion of the total power is emitted at higher frequencies. Both spectrograms correspond to chunk identified by $t^\star=130.5$ days, but with duration of 10 (top) and 5 (bottom) days, respectively. 
    Dashed lines denote boundaries of a chirp slice, as defined in~\cref{subsec:detection_strategy}: in particular, we show the slice matching the MBHB-3 signal as an outlier in our detection statistics. In the default configuration, the chirping tracks of MBHB-3 and MBHB-4 are clearly visible, whereas MBHB-2 is only marginally. By contrast, the former is clearly visible in the dedicated high-frequency configuration (bottom).}
    \label{fig:Figure3}
\end{figure}
Being located relatively nearby and far from the Galactic center (at a distance of $0.1~{\rm kpc}$, Ecliptic longitude $l=139^\circ$ and latitude $\beta= 4^\circ$) its signal could potentially be missed by noise-estimation routines -- typically targeting the slowly evolving power from the Galactic center -- and identified only later by high-latency pipelines.

Each data spectrogram (which we identify for convenience by its last timestamp $t^\star$)  is then whitened using the background PSDs
\begin{align}
\widehat{\mathrm{PSD}}[p,q] &= \frac{\mathrm{PSD}[p,q]}{\mathrm{PSD}_0[q;t_{\rm u}]}\,,
\label{eq:stftpower-whitened} 
\end{align}
where $t_u$ is chosen to be the unique one satisfying $10~{\rm d}<t^\star - t_{\rm u}<10.71~{\rm d}$ for a given $t^\star$, i.e. the closest preceding the data chunk. 
At runtime, multiple algorithms will run concurrently to identify transients, including MBHB mergers and instrumental artifacts. Therefore, the chosen delay of approximately 10 days provides sufficient margin to identify and mitigate unwanted transients that could otherwise contaminate the background estimation.

In~\cref{fig:Figure3} we show two examples of whitened spectrograms, obtained from data segments of duration 5 and 10 days: as we shall see in~\cref{subsec:detection_strategy}, this is the fundamental datum for our detection strategy.

\subsection{Detection}
\label{subsec:detection_strategy}

Our detection algorithm is based on searching for excess power in spectrograms with the morphology of a chirping signal. 
We adopt a simplified approach and model MBHB signals considering only the quadrupole mode.
In this framework, a \textit{chirp track} in the time–frequency plane is determined solely by the detector-frame chirp mass $\mc$ and time-to-coalescence $\tc-t$~\cite{maggioreGravitationalWavesVol2007}:
\begin{equation}
   f(t; \mc,\tc) = \frac{1}{\pi} \left( \frac{5}{256} \frac{1}{\tc - t} \right)^{3/8} \left( \frac{G \mc}{c^3} \right)^{-5/8}.
   \label{eq:chirp_track}
\end{equation} 

In this simple model, we neglect the time and frequency dependency of the detector response, effectively working in the \textit{frozen low-frequency} approximation~\cite{2021PhRvD.103h3011M}. 
This is justified by the data chunk duration of 10 days, short enough with respect to the detector motion timescale at frequencies below $5\times10^{-3}~{\rm Hz}$.
While to define a time-frequency correspondence we consider only the fundamental $(2,2)$-mode, the MBHB signals in SangriaHM are generated simulating $(2,2), (2,1), (3,3), (3,2), (4,3)$ and $(4,4)$ harmonics. 
The impact of higher-order modes depends {}{} on the mass ratio {}{} of the binary system and the inclination angle $\iota_0$ between the orbital angular momentum and the observer. Sources in the SangriaHM dataset are circular and non-precessing, with mass ratios spanning the range $0.36 < q < 0.97$, and inclination angles in the range $30^{\circ} < \iota_0 < 90^{\circ}$. The results presented in~\cref{sec:results} demonstrate that for these systems, a quadrupole-only model is sufficient to detect the signals and derive approximately correct constraints on $(\mathcal{M}_c, t_c)$.

We define a uniform grid in $\log_{10}\mc$ and $t_c^\star =\tc -t^\star$ where, for convenience of notation, we shift the time-of-coalescence by $t^\star$.
We further define a \textit{chirp slice} as the set of pixels
$\mathfrak{s} = \lbrace p_i,q_i, i=1,\dots,\Ns\rbrace$
whose centers in the TF plane are enclosed between two chirp tracks. 
We label each slice by a pair of values $(\log_{10}\mathcal{M}_{\mathrm{c}}, t^{\star}_{\mathrm{c}})$ defined as the arithmetic mean over 
those defining the two bounding chirp tracks.
To better map signals into the $(\mc, \tc^\star)$ space, adjacent slices are shifted so as to maintain a significant overlap.
In the default configuration, we limit our search to systems with $2\times 10^5 M_\odot<\mc<5\times 10^6 M_\odot$, and $-1 {\rm d} <t_c^\star<20 \,{\rm d} $, resulting in chirp slices defined by
\begin{align}
\log_{10}(\mc^{(n)}/{M_\odot}) &= \log_{10}(2\times 10^5) + 0.012837\times n \label{eq:gridmc}\,,\\
n & =0,\ldots,104 \,, \notag\\
\tc^{\star (m)} &= -1{\rm d} + 12000~{\rm s} \times m\,,\label{eq:gridtc}\\
m & =0,\ldots,148 \,.\notag
\end{align}
Although the search ranges for $\mathcal{M}_c$ and $t_c^\star$ were tailored to the SangriaHM dataset, they can be readily extended. 
While the computational cost scales linearly with the number of grid points in a serial execution, the analysis of each slice is entirely independent and therefore trivial to parallelize.
We only consider slices containing more than 10 TF pixels.

As shown in~\cref{fig:Figure3}, the proposed slices naturally accommodate for pixel power from merging MBHB signals.
Depending on the chosen STFT window length and the STFT hopping window~\cite{2018arXiv180707797R}, signals do not appear as thin lines, rather as thick bands in the TF plane: 
a longer window yields higher frequency resolution, making the low-frequency inspiral sharper and spreading out the rapidly chirping portion of the signal. 
In this case, the signal can be bracketed by two chirp tracks with relatively close $\mc$ values and widely separated $\tc^\star$ values. 
We show the signal of MBHB-3 as an example of this behaviour in the top panel of~\cref{fig:Figure3}, where the STFT window length is $\sim1.15$ days.
Conversely, a shorter window reduces the frequency resolution, broadening the inspiral and sharpening the high-frequency chirp power. 
Therefore, the same signal is enclosed by tracks with more widely separated $\mc$ and closer $\tc^\star$. 
We show in the bottom panel of~\cref{fig:Figure3} this configuration, where we set the STFT window length to $\sim1.39$ hours.

We proceed with our detection algorithm tiling the TF plane with overlapping chirp slices and test each slice for excess power.
Following the procedure described in~\cref{subsec:tf_representation}, we first identify the whitening ${\rm PSD}_0$ from the closest $t_{\rm u}$ to $t^\star-10 {\rm d}$. We then whiten the chunk spectrogram and compute the average power in each slice
\begin{equation}
\label{eq:average_asd}
\psd = \frac{1}{\Ns}\sum_{i=1}^{\Ns} \widehat{\mathrm{PSD}}[p_i, q_i]\,,
\end{equation}
where the sum runs over the $\Ns$ TF pixels selected by the slice $\mathfrak{s}$.

\begin{figure}[t]
    \centering
    \includegraphics[width=1\columnwidth]{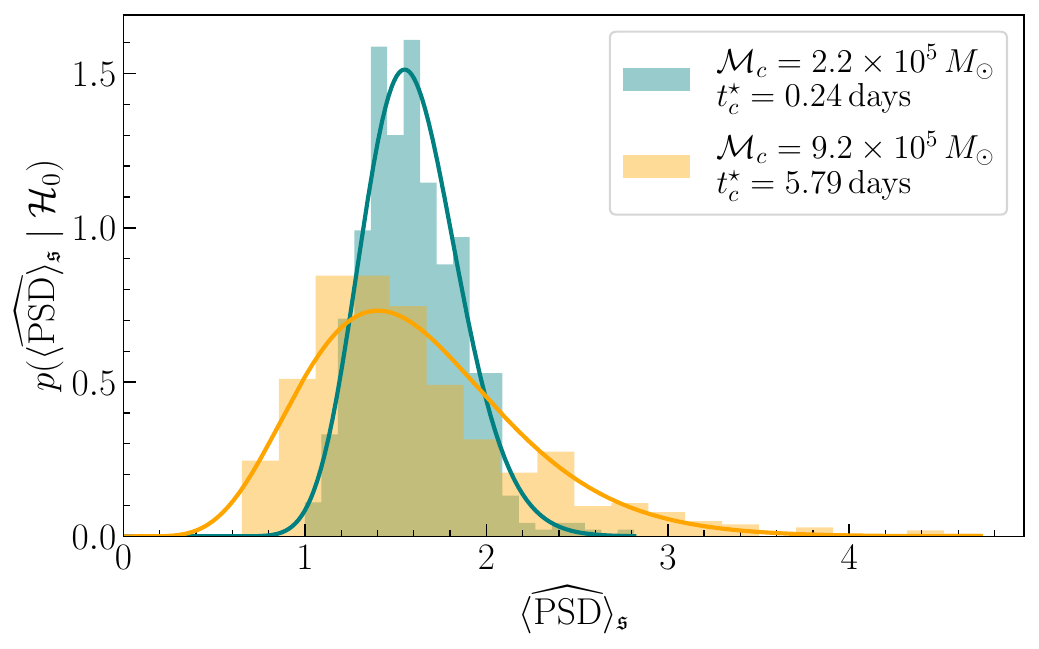}
    \caption{Example of background distributions relative to two spectrogram slices, in orange and teal, respectively. The distribution shape is affected by number of time-frequency points selected by each slice. 
    Supported by the derivation in~\cref{app:background_distro_derivation}, we find a Gamma distribution to be an adequate parametric family to fit for the observed distribution, obtained as described in~\cref{subsec:detection_strategy}. Shaded histograms and solid curve denote the synthetic realizations and the best fit obtained following the model in~\cref{eq:gamma_model}, respectively.
    }
    \label{fig:Figure4}
\end{figure}
\begin{figure*}[t]
    \includegraphics[width=2\columnwidth]{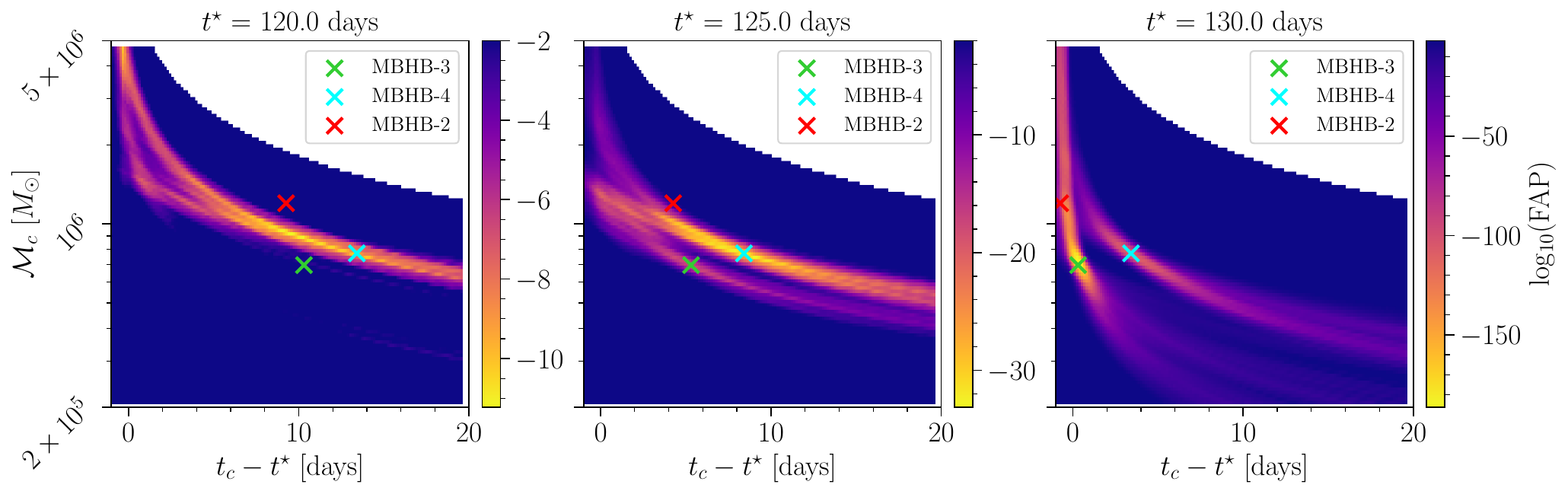}
    \caption{FAP maps at different chunk times $t^\star$, following the default search configuration described in~\cref{subsec:detection_strategy}. The white regions in the top-right corners correspond to $(\mc, \tc^\star)$ values associated with slices of the time–frequency plane that contain no spectrogram points because they fall outside its boundaries. We also remove the thin region corresponding to slices containing fewer than 10 points. The markers show the true chirp mass and coalescence time from the SangriaHM dataset metadata. 
    As the chunk timestamp approaches the coalescences, two regions in the $(\mathcal{M}_{\mathrm{c}}, t^{\star}_{\mathrm{c}})$ space emerge and gradually concentrate around the true values. 
    As discussed in~\cref{sec:results} in the default configuration, MBHB-2 is not detected by our algorithm.
    }
    \label{fig:Figure5}
\end{figure*}

To quantify the significance of a given slice averaged power, we compare the value from Eq.~\eqref{eq:average_asd} against a background distribution derived under the noise hypothesis $\mathcal{H}_0$ (i.e., in absence of MBHB signals). 
For this purpose, we use the SangriaHM dataset containing only instrumental noise and Galactic sources. 
To collect background realizations, we construct 500 whitened spectrograms equally spaced across one year, following the same procedure described in~\cref{subsec:tf_representation}.
Since the background PSD estimates used for whitening the spectrograms include the cyclostationary component of the Galactic DWD foreground, the collection of whitened background spectrograms can be approximately treated as independent realizations of a stationary stochastic process.

Once a set of chirp slices is defined in the $(\mc,\tc^\star)$ space, we collect for each slice 500 samples characterizing its $\psd$ distribution, reflecting both the number of TF pixels and the range of frequencies spanned. 
We find that a Gamma distribution provides an accurate two-parameters model across all slices, parameterized by $(a,b)$:
\begin{equation}
f(x;a,b) = \frac{(bx)^{a-1} e^{-(bx)}}{b\,\Gamma(a)}.\label{eq:gamma_model}
\end{equation}
In~\cref{fig:Figure4} we show two examples of fitted distributions for slices with different $(\mc,\tc^{\star})$ values. 
Ideally, one would need independent realizations of the background process to collect $\psd$ samples. 
However, since the 500 spectrograms are obtained from 10-days-long chunks equally distributed over one year, they overlap by approximately 93\%.
This introduces correlations across pixel power in the definition of the background distributions, especially for chirp slices spanning longer time intervals in the TF plane (lower $\mc$ values).

To justify the choice of the Gamma distribution, in~\cref{app:background_distro_derivation} we provide an analytical derivation of the background $\langle {\rm PSD} \rangle_\mathfrak{s}$ distribution under a set of simplifying assumptions. We demonstrate that the random variable $\langle {\rm PSD} \rangle$, computed over a set of uncorrelated time-frequency pixels, follows a Gamma distribution. We then extend this derivation to account for correlated pixels (arising from overlapping STFT windows) and obtain an explicit expression for the survival function. 
These results support the use of a Gamma distribution to model the background $\psd$. However, we maintain the use of a parametric fit, as it provides a more flexible approach that can account for imperfections in the whitening process.
The interested reader may find a general derivation in the context of stochastic gravitational wave backgrounds in~\cite{Baghi:2026dfk}.

Given the fitted background probability density $p(\psd\mid {\cal H}_0)$ for each slice, we quantify the false alarm probability (FAP) as the survival function of $\psd$ observed from data:
\begin{align}
\text{FAP}(x) &= \mathrm{Prob}( \psd > x \mid {\cal H}_0) \nonumber \\
&= 1 - \int_{0}^{x} p_{\psd}(u \mid {\cal H}_0) \, du.
\label{eq:survival_function}
\end{align}

For each chunk, we obtain a FAP \textit{map}, containing the FAP value relative to every slice $\mathfrak{s}$, as defined by the two values $(\mathcal{M}_{\mathrm{c}}, t^{\star}_{\mathrm{c}})$.
\Cref{fig:Figure5} illustrates an example of such maps, obtained from data containing the signals of MBHB-2, 3, and 4 for different chunk timestamps $t^\star$.
A threshold on the FAP is then applied to identify significant outliers.
In our default configuration, the significance threshold is set to $\faptr = 10^{-6}$. In the following section we describe the collection and processing of triggers, and justify the choice of the FAP threshold value.

\subsection{\label{subsec:coherent_refinement} Source identification and triggers tracking}

Once the FAP maps in the $(\mc, \tc^\star)$ parameter space are produced by the detection algorithm, we obtain parameter estimates and uncertainties from them.
To this end, we implement a labelling algorithm that identifies \emph{connected regions}: collection of adjacent slices, i.e. connected points in the FAP map domain.
We progressively scan increasing FAP values, moving upward from the global minimum, and identify connected components using feature labelling~\cite{labelling}, as implemented in \texttt{scipy.ndimage.label}~\cite{Virtanen:2019joe}.
Climbing up higher FAP levels, initially disconnected regions merge, as saddle points between local minima are crossed. 
The algorithm monitors the number of distinct regions at each threshold: 
a reduction in the number of regions indicates that the threshold has crossed a saddle point between two previously independent minima.
The procedure continues up to a predefined maximum FAP level.

To limit the effect of noise fluctuations --particularly at the high FAP levels typical of early-detections -- we apply some quality cuts when identifying regions: (i) we enforce a minimum size of 30 connected grid points, and (ii) we automatically merge distinct regions whose local minima are separated by fewer than 15 grid points. 
These cuts are empirically tuned in the default configuration to suppress the regions fragmentation observed for early source identification in the SangriaHM dataset.
Each identified region is then assigned a best estimate of $(\mc, \tc^\star)$, by its point of minimum FAP. Uncertainties can then be defined in several ways, depending on the analysis requirements.
In this work, we adopt for each FAP region contour levels corresponding to fixed $\log_{10}{\rm {FAP}}$ fractions of the local minimum.

\begin{figure*}
\begin{tikzpicture}[
    node distance=12mm and 18mm,
    every node/.style={font=\small},
    process/.style={
        rectangle,
        rounded corners,
        draw=teal,
        align=center,
        text width=6cm,
        minimum height=8mm
    },
    decision/.style={
        rectangle,
        draw=orange,
        align=center,
        text width=3.5cm,
        aspect=2,
        fill=orange!20
    },
    teal/.style={
        process,
        text=black,
        fill=teal!30
    },
    arrow/.style={
        ->,
        >=Stealth,
        thick
    }
]

\node[teal] (load) {
    \textbf{Load FAP map}\\
        \begin{enumerate}\setlength\itemsep{1pt}
            \item[i.] Apply ${\rm FAP}_{\rm max}$ threshold
        \end{enumerate}
};

\node[teal, right=of load,xshift=-1cm] (identify) {
    \textbf{Identify connected regions}
    \begin{enumerate}\setlength\itemsep{1pt}
        \item[i.] FAP level scanning
        \item[ii.] Identify connected regions
        \item[iii.] Detect saddle points
        \item[iv.] Define region boundaries
    \end{enumerate}
};

\node[teal, below=of identify] (filters) {
    \textbf{Apply region quality cuts:}
    \begin{enumerate}\setlength\itemsep{1pt}
        \item[i.] Minimum region size
        \item[ii.] Merge nearby regions
    \end{enumerate}
};

\node[decision, left=of filters] (exist) {
    Are there live\\ regions from past timestamps?
};

\node[teal, below=of exist] (match) {
    \textbf{Match with live regions}\\
    Loop over live regions:
    \begin{enumerate}\setlength\itemsep{1pt}
        \item[i.] Check overlap
        \item[ii.] Assign best match \\to largest overlap
    \end{enumerate}
};

\node[decision, right=of match] (found) {
    Found matching\\ region?
};

\node[teal, below=of found] (update) {
    \textbf{Update existing region}
        \begin{enumerate}\setlength\itemsep{1pt}
        \item[i.] Select points: union or intersection
        \item[ii.] Keep region label
    \end{enumerate}
};

\node[teal, below left=of found] (create) {
    \textbf{Create new regions}
    \begin{itemize}\setlength\itemsep{1pt}
        \item[i.] Assign new label
        \item[ii.] Set birth to $t^\star$
    \end{itemize}
};

\node[process, below=of update] (mergeold) {
    \textbf{Many-to-one overlap}
    \begin{enumerate}\setlength\itemsep{1pt}
        \item[i.] Keep old region with lowest FAP
        \item[ii.] Merge overlapping points
        \item[iii.] Deactivate other old regions
    \end{enumerate}
};

\node[process, below=of create] (mergenew) {
    \textbf{One-to-many overlap}
    \begin{enumerate}\setlength\itemsep{1pt}
        \item[i.] New region with largest overlap\\inherits old region label
        \item[ii.] Other new regions tracked separately
    \end{enumerate}
};

\node[teal, below=of mergenew] (save) {
    \textbf{Save regions}\\
    Deactivate regions \\unupdated for over two days
};

\draw[arrow] (load) -- (identify);
\draw[arrow] (identify) -- (filters);
\draw[arrow] (filters) -- (exist);

\draw[arrow] (exist) -- node[right]{YES} (match);
\draw[arrow] 
    (exist.west) -- ++(-1.6,0) 
    |- node[pos=0.25,left]{NO} (create.west);

\draw[arrow] (match) -- (found);

\draw[arrow] (found.south) ++(-0.4,0) |- ++(0,-0.9) -| node[pos=0.1,above]{NO} (create);
\draw[arrow] (found.south) --  node[pos=0.55,right]{YES} (update.north);

\draw[arrow] (update.south) ++(-0.4,0) |- ++(0,-0.9) -| (mergenew.north);
\draw[arrow] (update.south) ++(-0.0,0) |- ++(0,-0.9) -| (mergeold.north);

\draw[arrow] (create.south) ++(-2.0,0) |- ++(-0.8,-1.0) -| ++(-1.0,0) |- ++(0.0,-2.7) -- (save.north) ;

\draw[arrow] (mergeold.south) -- (save.north);
\draw[arrow] (mergenew.south) -- (save.north);

\draw[arrow]
     (save.west) -- ++(-2.2,0) |- (load.west)
     node[midway,above]{loop over chunk $t^\star$};

\end{tikzpicture}
\caption{Overview of region tracking scheme across data segments, as described in~\cref{subsec:coherent_refinement}.
Flowchart of the region detection and tracking algorithm applied to FAP maps over time. For each time chunk $t^\star$, the FAP map is thresholded, connected regions are identified, refined using quality cuts, and matched to previously active (``live'') regions based on overlap. Depending on the matching outcome, regions are newly created, updated (many-to-one), or split (one-to-many), with labels propagated accordingly. Regions that are not updated for more than two days are deactivated and saved.
}
    \label{fig:diagram2}
\end{figure*}
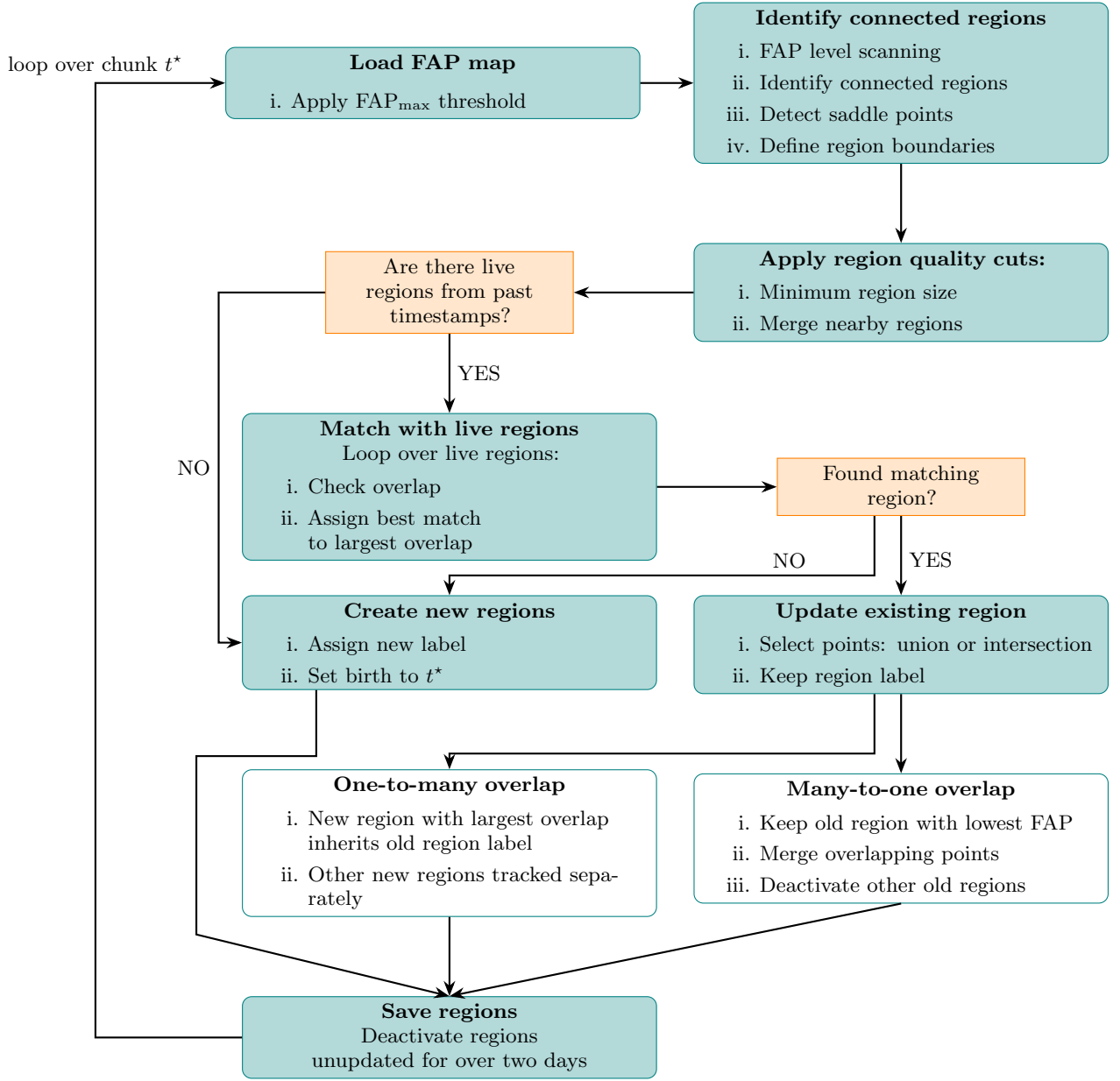

The shape of a signal-activated region in a probability map evolves over time, as illustrated in~\cref{fig:Figure5}. We track this evolution by matching regions identified at consecutive timestamps, allowing information to accumulate and progressively refine the parameter estimates. 
When a new region is detected, the algorithm checks whether it significantly overlaps (at least 10 grid points in common) with one or more region from a previous timestamp. 
If sufficient overlap is found, the new region is associated with the existing one. 
At this point, only the overlapping points are kept, and the region data are updated with the new information while preserving its original label.
When two previously tracked regions merge, due to a new one overlapping with both, the algorithm keeps only the overlapping grid points, labels them as the parent region with the lowest FAP, and marks the remainder as inactive. 
This behavior may be observed during the first stages of detection, when multiple local minima typically appear before merging into a single region.

On the other hand, when a tracked region splits into multiple ones at a subsequent timestamp, the new region with the largest overlap inherits the previous label, and the other new region is tracked independently. This behavior is observed when two overlapping signals become separable, like the case of MBHB-3 and 4.

Thresholds on the number of points per region and on the number of overlapping points are chosen empirically and depend on the grid density. These thresholds serve as temporary countermeasures to mitigate unwanted behavior, such as consecutive region-splitting and merging. 
In future work, a more robust control criteria based on the significance of local minima relative to the background FAP distribution, will be implemented.
In~\cref{fig:diagram2} we summarize diagrammatically the region-tracking scheme as just described.

In~\cref{fig:Figure6}, we instead illustrate the temporal evolution of the MBHB-0 parameter estimates, shown as relative error in chirp mass $\Delta\mc/\mc$ and difference in coalescence time $\Delta\tc$ with respect to the true injected values. 
The estimates converge toward the injected ones, reported in the SangriaHM dataset metadata; however, a small residual bias in the chirp mass ($\gtrsim2\%$) 
and coalescence time (a few hours)
may persist up to merger.

The current implementation of our algorithm enforces the retention of only overlapping points in the $\mc$-$\tc$ space during region tracking across successive timestamps. This choice ensures that constraints on $\mc$ established at the onset of tracking are preserved over time. However, an alternative version of the algorithm can be implemented without this requirement. In this case, the shape of the regions tracked at each timestamp in the $\mc$–$\tc$ space is no longer influenced by their past evolution. This approach can help prevent the loss of region tracking when early estimates are biased, but it may also increase confusion in the presence of overlapping sources. An example of this behavior is discussed in~\Cref{app:estimates_all_sangria} and illustrated in~\cref{fig:FigureA3}.

The choice of the FAP detection threshold determines the algorithm’s sensitivity to noise fluctuations. To quantify this effect, we evaluate the number of spurious regions detected when the algorithm is applied to data containing only instrumental noise and Galactic DWDs. Using a threshold of $\faptr = 10^{-6}$, which we adopt as our default value, the algorithm reports 10 false detections over one year of simulated data{} running on channel A, and 9 on channel E. False detections arise when background fluctuations cause the $\psd$ in certain time-frequency slices to exceed the chosen threshold, leading the tracking algorithm to identify spurious regions. The A and E TDI channels are noise-orthogonal, and we verify that all occurrences of false positives are not concurrent across them. The absence of a concurrent trigger in both channels could be used, in principle, as a smoking gun to identify noise-driven spurious detections. Particular attention should be paid, however, as even genuine GW signals can exhibit unequal power in the $A$ and $E$ channels, depending on the source extrinsic parameters (as reported in Sec.~\ref{sec:results}). We leave a characterization of multi-channel analysis to strengthen detections and remove false positives to a future version of the algorithm. {}
Setting the value of $\faptr$ entails a trade-off between the number of false detections and the earliest time at which sources can be identified. The optimal choice depends on whether we prioritize a conservative search with fewer false positives or aim to maximize our sensitivity to faint signals.

We apply this detection strategy in parallel to the TDI channels A and E. The response of the two channels to a given source differs depending on its position relative to the detector, as encoded in their antenna pattern functions~\cite{Tinto:2020fcc}. 
In current analysis, searches in the two channels are conducted independently. The activated regions in the resulting FAP maps are then compared, and a flag is raised when overlapping
regions are identified in both channels, following the same set of rules~described~in~\cref{subsec:coherent_refinement}.

\section{\label{sec:results} Results}

\begin{figure}[t]
    \centering
    \includegraphics[width=1\columnwidth]{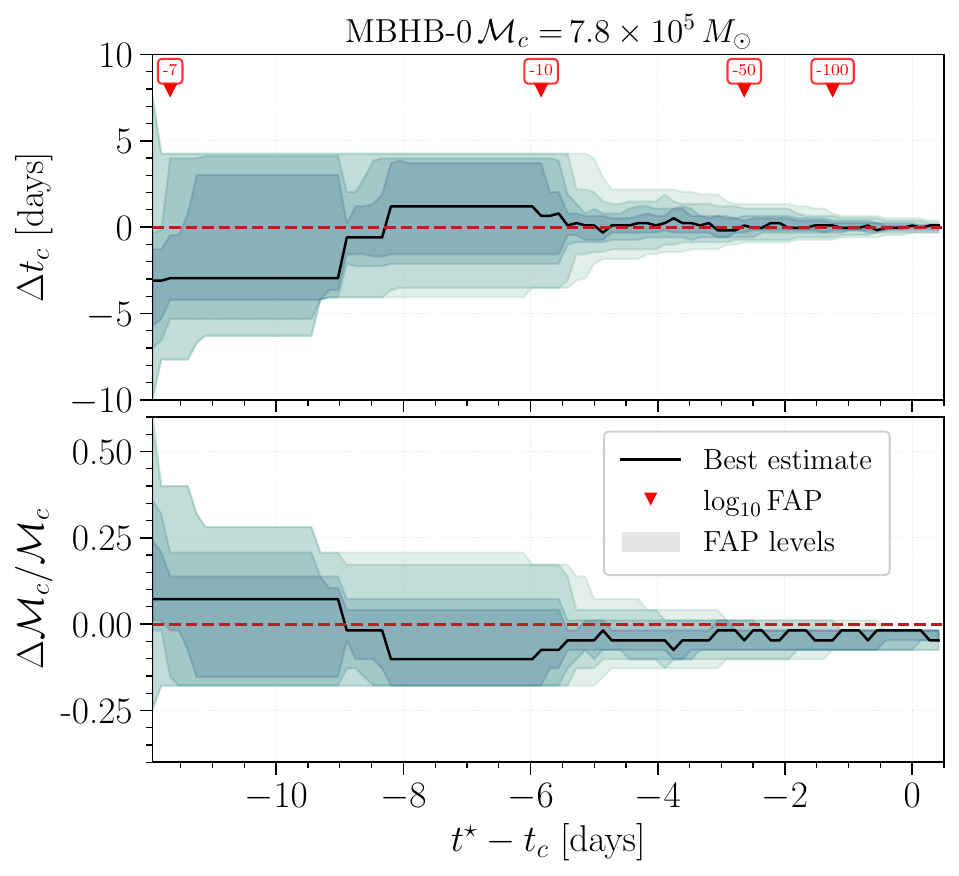}
    \caption{Evolution of $\mc$ and $\tc$ estimates for the signal MBHB-0, shown in top and bottom panel, respectively.
    Results are shown relative to the true injection, hence the red horizontal lines denote them.
    The black solid lines denote the best estimate, corresponding to the minimum FAP value, at each $t^\star$. 
    Shaded regions denote the point estimates uncertainties, as determined by the region tracking scheme described in~\cref{subsec:coherent_refinement}: decreasingly lighter shades correspond to regions where $\log_{10}{\rm FAP}$ is within $95\%$, $90\%$, $70\%$ and $50\%$ of the minimum value, respectively.
    For reference, a few red markers label the minimum FAP values achieved at corresponding $t^\star$. 
    }
    \label{fig:Figure6}
\end{figure}

The algorithm successfully identifies all 15 massive black hole binaries included in the SangriaHM dataset. \Cref{fig:Figure6} illustrate the evolution of the $\tc$ and $\mc$ estimates for the MBHB-0 signal, while in~\cref{app:estimates_all_sangria} we systematically illustrate results for all sources in the SangriaHM dataset. 

\begin{figure*}[t]
    \includegraphics[width=2\columnwidth]{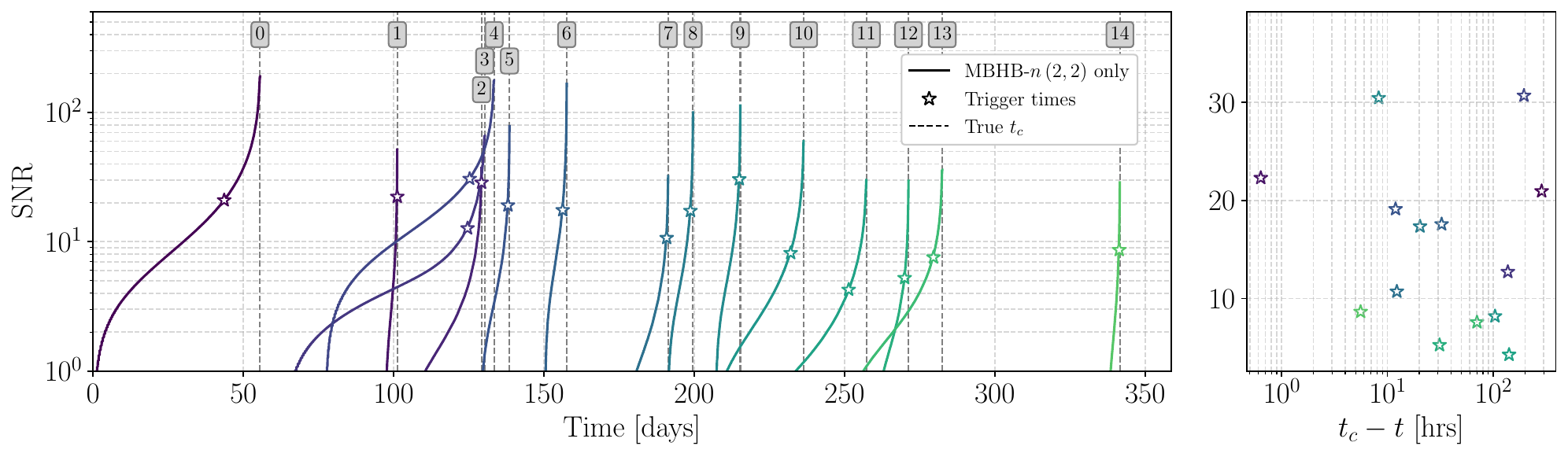}
    \caption{
    Evolution of SNR for every source in the SangriaHM dataset. 
    \textit{(Left panel)} Cumulative SNR for each source, as the mission progresses through the simulated year of data. The computation is carried out adapting the code base in \citet{Spadaro:2024tve}. We employ the IMRPhenomXHM approximant~\cite{Garcia-Quiros:2020qpx} to produce waveforms analogous to the signals in SangriaHM, along with an analytical model of the instrumental and DWD confusion noise. Individual signals, numbered by grey boxes, are simulated up to their time-of-coalescence (dashed grey lines), and a star marker denotes the time of first detection by our algorithm.
    \textit{(Right panel)} SNR value at the time of first detection, for a threshold $\faptr=10^{-6}$.
    The first-detection timestamps shown in the figure are consistent with the results presented in~\cref{fig:FigureA1,fig:FigureA2}: the default algorithm configuration is used for the majority of signals, while MBHB-1 and MBHB-2 are analyzed using a time–frequency configuration with a shorter STFT window, as described in~\cref{subsec:detection_strategy}.
    As discussed in~\cref{sec:results}, MBHB-2 is identified $1.75$ hours after coalescence, hence is not shown in the right panel.
    }
    \label{fig:Figure7}
\end{figure*}

Detection and tracking performance vary across the 15 sources. 
All binaries but MBHB-2\footnote{MBHB-2 requires an adaptation of the algorithm configuration, discussed further in the text.} are detected before merger, with MBHB-0, 3, 4, 10, 11, and 13 identified between 2 and 17 days in advance. For these sources, the best estimates for $\mc$ and $\tc$ overlap with the true values with an accuracy up to two grid steps a few days before merger.
In the case of premerger detections, FAP minima typically approach the true values as more data are accumulated. 
Signals with a more prominent inspiral phase yield the highest precision on $\mc$ estimates at merger time\footnote{The estimates of $\mc$ and $\tc^\star$ continue to evolve after the merger time $\tc$ as additional data become available, since the spectrograms do not extend up to the current timestamp $t^\star$, as described in~\cref{subsec:tf_representation}.}, with relative errors below $3\%$ (MBHB-0, 3, 4, 10, 13).
MBHB-7 and 11 exhibit fractional uncertainties between $4\%$ and $7\%$, while signals with weaker inspiral--such as MBHB-2, 5, 6, 8, and 14--show uncertainties ranging from $20\%$ to $40\%$. 
Finally, MBHB-1 is recovered with the highest uncertainty, reaching a relative error of $138\%$. 
These discrepancies may arise from the strong approximations we make in signal modeling. 
However, we do not consider this a critical issue, as the goal of our method is to provide a rapid and approximate source parameters estimate, to serve as prior constraints to downstream algorithms performing more accurate parameter estimation.

Depending on their parameters--most notably the detector frame masses--MBHBs enter the LISA band and accumulate signal-to-noise ratio (SNR) at different frequencies. In~\cref{fig:Figure7}, we show the evolution of the cumulative SNR for all sources and highlight the SNR values at the time of first detection. 
Signals such as MBHB-1, MBHB-2, and MBHB-14 have large chirp masses ($\mc > 10^{6} M_{\odot}$). For these systems, the accumulated SNR rises steeply, and detection occurs only close to, or slightly after, the merger.
Adapting the configuration of the time–frequency representation can be beneficial for the detection of such signals. 
In the case of MBHB-1 and MBHB-2, we decrease the frequency resolution from $10^{-5}\,\mathrm{Hz}$ to $2\times10^{-4}\,\mathrm{Hz}$, consequently increasing the time resolution and reducing spectral leakage at higher frequencies, thereby enhancing the late-inspiral 
part of the
signal. This strategy is described in~\cref{subsec:detection_strategy} and illustrated in the bottom panel of~\cref{fig:Figure3}. The results shown in~\Cref{fig:FigureA1,fig:FigureA2} relative to these signals are obtained using such dedicated configuration.

\begin{figure}[t]
    \centering
    \includegraphics[width=1\columnwidth]{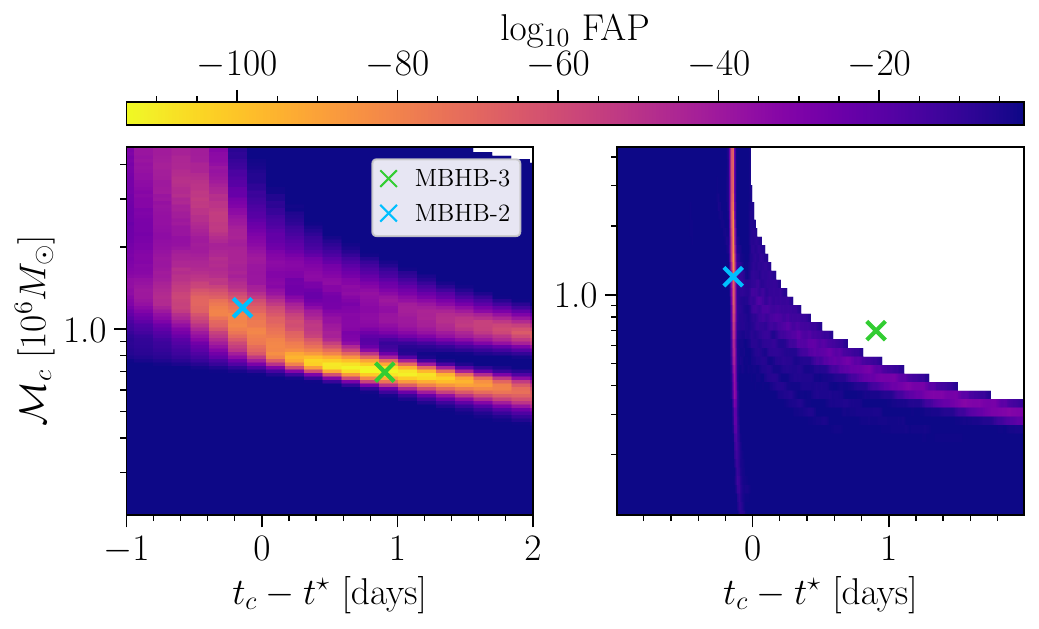}
    \caption{Comparison of FAP maps for the same data chunk under different time–frequency configurations. The left panel shows the default settings, while the right panel uses a higher-frequency band with lower frequency resolution. In the left panel, the regions activated by MBHB-3 and MBHB-4 are clearly visible, but no distinct peak appears at the location of MBHB-2. In contrast, the FAP map on the right displays a discernible low-FAP region corresponding to MBHB-2.
    }
    \label{fig:Figure8}
\end{figure}

With our default configuration, MBHB-2 is not detected, as its signal overlaps significantly with the stronger emissions of MBHB-3 and MBHB-4, particularly at lower frequencies.
We identified a strategy to mitigate such overlap in time–frequency by adjusting the spectrogram frequencies. 
Our default setup spans frequencies between $10^{-4}\,\mathrm{Hz}$ and $10^{-3}\,\mathrm{Hz}$. 
By shifting to a higher-frequency band ($5.0\times10^{-4}\,\mathrm{Hz} < f < 5.0\times10^{-3}\,\mathrm{Hz}$), we effectively suppress the loud inspiral portions of MBHB-3 and MBHB-4. 
Moreover, as discussed above, we reduce the frequency resolution at the advantage of an increased time resolution. In~\cref{fig:Figure8} we show the FAP maps obtained for these different configurations, demonstrating that MBHB-2 becomes clearly separable from  other signals and is successfully detected. 
However, the signal is correctly detected $1.75$ hours after coalescence. 
Although a region of the parameter space exceeds the detection threshold before merger, the corresponding parameter estimates are significantly biased, and the region is not tracked further due to the lack of overlapping points at subsequent timestamps.
This demonstrates that running multiple searches in parallel, each optimized for different frequency ranges, may be necessary during operations to address source overlap. 
The minimal numerical burden of our search allows the adoption of such strategy remaining well below a minute, i.e. the envisioned low-latency analysis timescales.

For each source in the dataset, we additionally compare the evolution of $\min \log_{10}\mathrm{FAP}_{\mathrm{A,E}}$, i.e. the FAP minima as recorded by the region-tracking algorithm in the two TDI channels.
Depending on the source parameters, their ratio typically spans up to two orders of magnitude, with $0.1 < \min \log_{10}\mathrm{FAP}_{\mathrm{A}} / \min \log_{10}\mathrm{FAP}_{\mathrm{E}} < 10$.
For MBHB-1, MBHB-2, and MBHB-13, the measured power in channel E is consistently higher than in channel A throughout the tracking. The opposite behavior is observed for MBHB-6, MBHB-7, and MBHB-10, while MBHB-5 is not detected in channel E. The observed differences in measured power across channels lead to different first-detection times, with offsets of up to $\pm 4$ days. For all other signals, we measure a comparable amount of power in both channels.

We assessed the algorithm computational performance by measuring the runtime of its main components on a single core.
The workflow consists of two main stages: the computation of probability maps and the region-tracking  algorithm. 
Based on 200 repetitions, we recorded a mean computing time for a single spectrogram of $0.039\pm 0.004\,\mathrm{s}$. 
Generating probability maps is more demanding, with an average runtime of $0.43 \pm 0.03\,\mathrm{s}$ per map, i.e. the dominant computational cost of the algorithm.
The total processing time for the region-tracking algorithm varies depending on the number of regions being simultaneously tracked. The average runtime per chunk is $0.14\pm0.06\,\mathrm{s}$.

An issue identified during our analysis is the spurious activation of regions in the FAP maps caused by the loud signals of already merged sources. When the merger phase of an MBHB enters the spectrogram, a large amount of power is concentrated in the high-frequency chirp portion. 
Slices of the $(\tc,\mc)$ grid intersecting such signal may exceed the detection threshold in $\psd$, even though they merely cross the signal rather than following its chirping evolution. 
These merger-induced activations appear in the FAP maps as new regions characterized by low chirp masses and may be mistakenly interpreted as distinct sources. 
For instance, following the merger of MBHB-0, approximately 20 spurious regions are activated.
This issue can be mitigated through a masking strategy. Once a source is confidently identified (e.g., $\fap \approx 10^{-10}$), we remove the time–frequency pixels associated with slices in a small neighborhood around the minimum-$\fap$ point of the tracked region. 
This procedure effectively removes the tracked signal from the time–frequency plane, while leaving the computation of $\psd$ over other slices unaffected. 
We tested this approach and verified that such spurious regions are efficiently suppressed.

Spurious region activation may also occur in other situations. 
In rare cases, a tracked region may fragment due to statistical fluctuations, temporarily producing two nearby regions; however, the spurious region is typically short-lived and disappears within a few hours. In the presence of multiple overlapping signals, their interaction may also generate spurious regions in the FAP map. This is illustrated by the case of MBHB-3 and MBHB-4, which have similar chirp masses: their corresponding regions become clearly separable only about 5 days before the merger of MBHB-3, during which approximately 15 spurious regions are activated. 
We verified that these can again be mitigated using the masking strategy described above. 
Since MBHB-4 is the loudest signal in the dataset, it is detected before MBHB-3 despite merging later. 
Applying the masking to MBHB-4 safely removes both its signal and the spurious regions arising from its confusion with MBHB-3, enabling the detection of MBHB-3 up to 8 days before merger.

Finally, we offer a few considerations regarding glitches. In the time–frequency domain, known glitches appear as short-lived (minutes to hours), high-amplitude transients spanning a broad frequency range, potentially mimicking the merger phase of an MBHB coalescence~\cite{Spadaro:2023muy}. Although a systematic study of this effect is deferred to future work, we expect that, in the current configuration, unflagged glitches could be misidentified as loud, high-mass signals, which are characterized by the rapid inspiral described above. Aside from this potential mislabeling, we anticipate the algorithm to be fairly robust: a glitch occurring during the early inspiral of an MBHB produces a clearly distinguishable time–frequency signature, and we have shown that the algorithm can handle overlapping sources in that regime.
 
\section{Conclusions}
\label{sec:conlusions}

In this work, we presented a fast prototype algorithm to detect MBHBs with LISA in low-latency. 
The main objective is to issue early alerts with approximate chirp mass and coalescence time estimates, potentially triggering protected observational periods and providing informative priors for more detailed parameter-estimation algorithms.

Our detection strategy is based on the search for excess power with the morphology of a chirping signal in time–frequency data. 
We compute the STFT of 10-day data segments and accumulate the power over pixels defined by the quadrupole frequency evolution. 
The algorithm outputs a significance map which measures the FAP of the average power inside each slice, relative to a background distribution. The background distribution is constructed from data segments containing only instrumental noise and Galactic sources, and is phenomenologically fitted with a Gamma function. 
At each timestep, the significance maps are reprocessed by a region-tracking algorithm: it targets points in the $(\mc, \tc)$ space exceeding a significance threshold, labels them as potential sources, and follows their temporal evolution while progressively refining the estimates of $\mc$ and $\tc$. 
Multiple potential sources can therefore be tracked simultaneously from a single chunk of data.
A complete analysis for a single 10-days-long chunk of data is typically complete in $0.4$ s using a single core.

For sources characterized by long inspirals, detected a few days before merger, parameter estimates yield typical fractional errors of 1–40\% in chirp mass, and a few hours in coalescence time.
The majority of MBHB sources in the SangriaHM dataset are detected premerger, typically hours to days before coalescence. The signal of a particularly massive and high-redshift source is detected only after merger, primarily due to overlap with two other signals exhibiting louder inspiral phases. Some signals required tuning of the time-frequency configuration for identification, as discussed in~\cref{sec:results}.
In a realistic scenario, such tuning would be replaced by global optimization over multiple simulated datasets, producing a set of search configurations to be run in parallel for a complete exploration of data.

In \citet{CabournDavies:2024hea}, the authors adapt several well-established methods from ground-based detectors to the LISA context for both the search and parameter estimation of MBHB merger signals. 
The search algorithm is based on matched filtering against an optimally populated template bank of premerger waveforms, allowing them to confidently detect signals with different morphologies once they reach SNR~$\sim 8$. The algorithm proposed was tested on a set of five representative MBHBs with $7.3\times10^5 M_\odot < \mc < 8.7 \times 10^6 M_\odot$ (detector frame), and $0.125<q<1$. 
In our approach, as shown in~\cref{fig:Figure7}, SangriaHM sources first detection does not occur at a fixed SNR threshold.
Instead they take place over a fairly broad range of values, approximately $4 \lesssim \mathrm{SNR} \lesssim 30$.

The results obtained for the sources in the SangriaHM dataset are consistent with those reported in \citet{Deng:2025qhx}, which 
presented the first application of a premerger detection algorithm on the Sangria dataset (which differs from SangriaHM by having only the (2,2) harmonic present in the signals of MBHBs). The two approaches are fundamentally different{} , as the method presented in \citet{Deng:2025qhx} is based on the maximization of the $\mathcal{F}$-statstic on the intrinsic parameters of the source. 
The algorithm in~\citet{Deng:2025qhx} is able to detect every source in the Sangria dataset previous to its merger. However, our algorithm provides useful preliminary estimates of the chirp mass and coalescence time with a computational cost roughly two orders of magnitude lower. 
In fact, the computational cost of the end-to-end analysis of single 10-day data chunk is currently dominated by FAP map generation ($\sim0.4$ s) and region tracking ($\sim0.14$ s). 
By contrast, the algorithm presented in \citet{Deng:2025qhx} currently requires $\sim15 {\rm min}$ to analyze 15 days of data, where the main computational cost resides in the waveform evaluation. 

In the current implementation, our algorithm is applied independently to $A$ and $E$ TDI channels. In future work we will investigate a detection statistic coherently combining information from multiple channels.

As discussed in~\cref{subsec:detection_strategy}, the background distributions are estimated under the idealized assumption that all transient signals occurring before the last 10 days of data have been detected and subtracted. It is foreseeable that an online estimate of the PSD will be provided by other blocks of the LISA ground-segment (e.g., the noise-estimation global fit blocks), possibly including an accurate characterization of the time-evolving confusion noise~\cite{Buscicchio:2025zeb}.

Alternative time–frequency representations could replace the STFT without altering the algorithm logic and structure. 
Notably, the constant-Q transform~\cite{2004CQGra..21S1809C, 2024PhRvD.109j2010V} have been shown to provide a multi-resolution time–frequency representation that can reduce the impact of spectral leakage effects. 
Preliminary analyses suggest that using the constant-Q transform may allow better constraints of $\tc$ and $\mc$. 
A systematic comparison between the two methods is left for future work.

The time–frequency domain offers a straightforward way to handle data gaps, as the flexible selection of time–frequency bins allows the isolation of those affected by such artifacts. We plan to implement an automated procedure to mitigate their impact on detection and preliminary parameter estimation.

Currently, the algorithm discards information from the amplitude and phase evolution of the signals, which may encode crucial information about source extrinsic parameters. Future developments incorporating the analysis of such features are under investigation for an improved source characterization including critical, additional source parameters like their sky-localization.

\begin{acknowledgments}
The authors are grateful to Q.~Baghi, F.~Pozzoli, A.~Spadaro, J.~B.~Bayle, C.~Cossou, R.~Srinivasan, L.~Copparoni, F.~Rigamonti, M.~Le~Jeune, S.~Deng, S.~Babak and D.~Bandopadhyay for discussions and valuable inputs.
R.B. is supported by the Italian Space Agency grant ``Phase B2/C activity for LISA mission'', Agreement n.2024-NAZ-0102/PE, MUR Grant ``Progetto Dipartimenti di Eccellenza 2023-2027'' (BiCoQ), the ICSC National Research Centre funded by NextGenerationEU.
Computational work was performed at CINECA with allocations through INFN and Bicocca.

\end{acknowledgments}

\newpage
\bibliographystyle{apsrev4-1}
\bibliography{main}

\newpage
\onecolumngrid
\appendix
\renewcommand{\thefigure}{A\arabic{figure}}
\renewcommand{\theHfigure}{A\arabic{figure}}
\setcounter{figure}{0}

\section{$\mathcal{M}_c$ and $t_c$ estimates across all Sangria sources}
\label{app:estimates_all_sangria}

\begin{figure*}[t]
    \includegraphics[width=0.9\columnwidth]{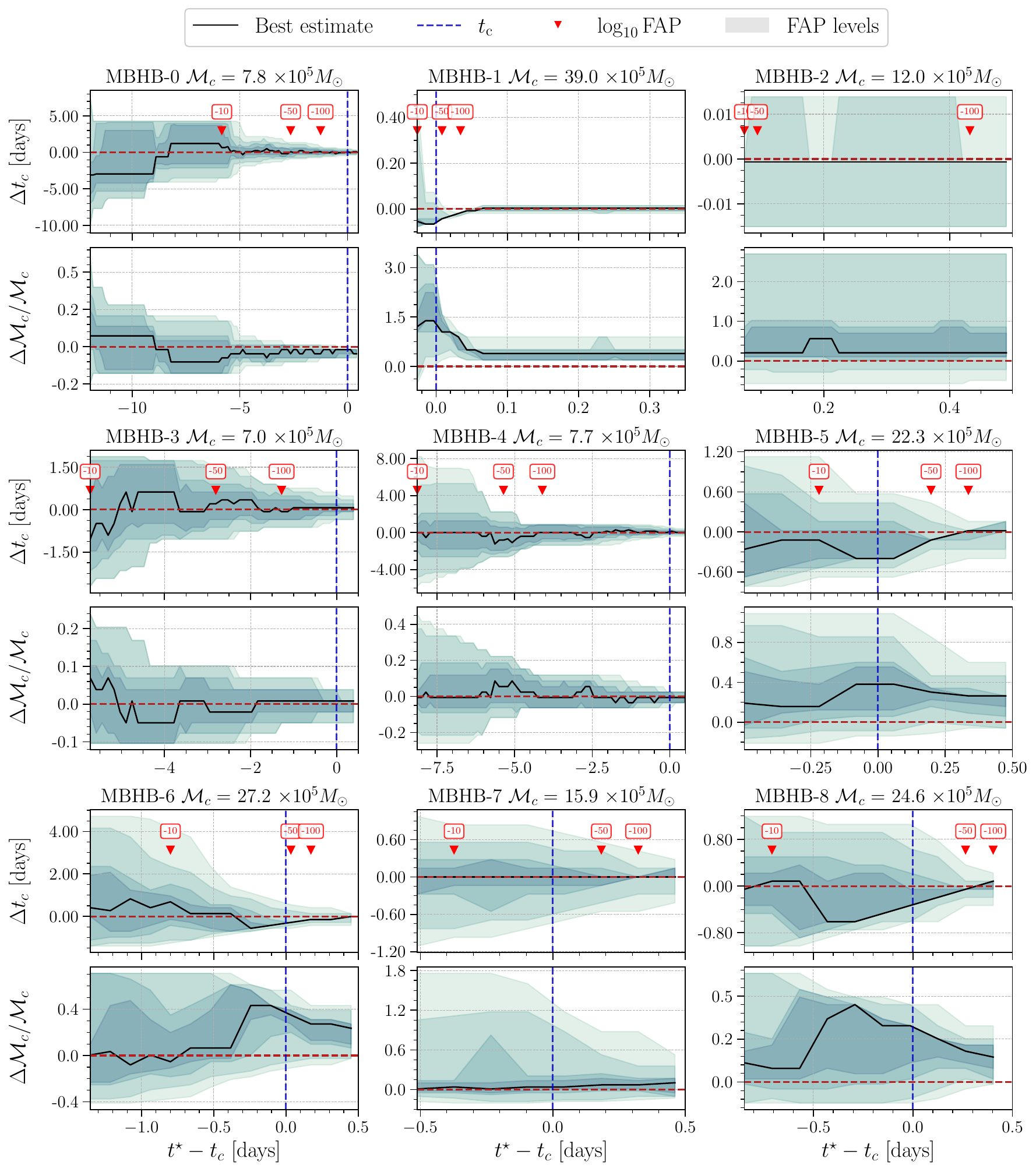}    
    \caption{Evolution of $\mc$ and $\tc$ estimates across the first 9 Sangria sources, as a function of the data processed up to $t^\star$. Each subplot corresponds to one source.
    Red horizontal lines in each subplot correspond to the true injected values, i.e. zero relative differences with respect to our estimates. 
    Blue vertical lines marks the merger time for each system, and black solid curves shows the best estimate at each $t^\star$.
    Shaded regions denote the point estimates uncertainties: decreasingly lighter shades correspond to regions where $\log_{10}{\rm FAP}$ is within $95\%$, $90\%$, $70\%$ and $50\%$ of the minimum value, respectively. 
    Parameter estimates for MBHB-2 are tracked from $1.75$ hours after coalescence, while all other signals are detected before merger. 
    Signals with a more prominent inspiral phase yield the highest precision on $\mc$ estimates at merger time ($\Delta \mc / \mc \lesssim 3\%$), while signals with weaker inspiral show uncertainties ranging from $20\%$ to $40\%$. The highest relative error is registered for MBHB-1 ($\Delta \mc / \mc =138\%$).
    Preliminary constraints of the coalescence time within a few hours with respect to the true value are generally obtained days before merger (hours in the case of signals characterized by shorter inspirals).}
    \label{fig:FigureA1}
\end{figure*}

\begin{figure*}[t]
    \includegraphics[width=0.9\columnwidth]{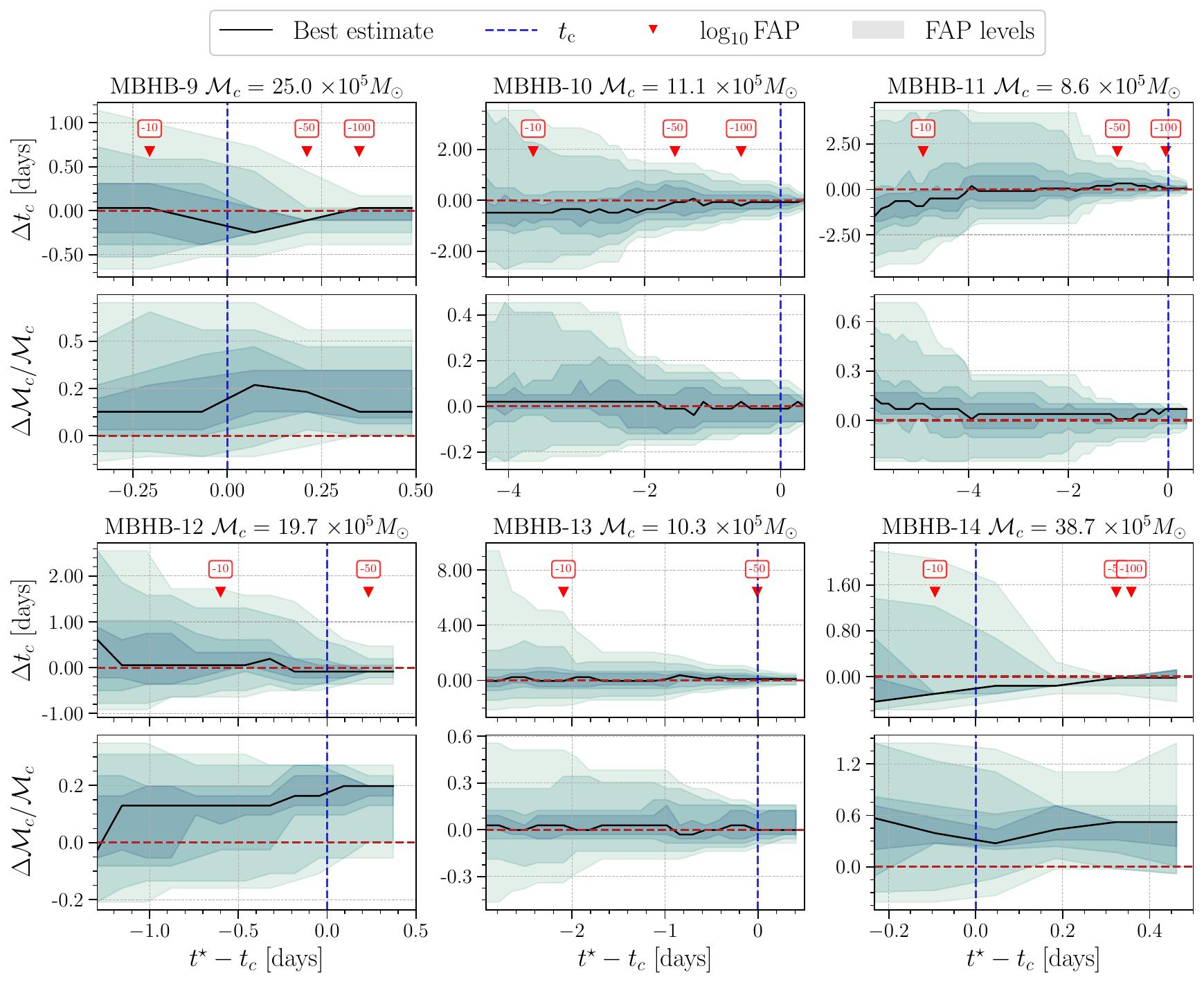}

    \caption{Evolution of $\mc$ and $\tc$ estimates across the last 6 Sangria sources, following the same conventions in~\cref{fig:FigureA1}.}
    \label{fig:FigureA2}
\end{figure*}

In~\cref{fig:FigureA1,fig:FigureA2}, we show the evolution of the $\mc$ and $\tc$ estimates for all Sangria sources.

The results shown for MBHB-1, MBHB-13, and MBHB-14 are obtained from the search performed using channel E, as the measured power is consistently higher than in channel A, as discussed in~\cref{subsec:coherent_refinement}. In channel A, sources MBHB-1 and MBHB-14 are detected only after merger.

\begin{figure*}[b]
    \centering
    \includegraphics[width=0.43\columnwidth]{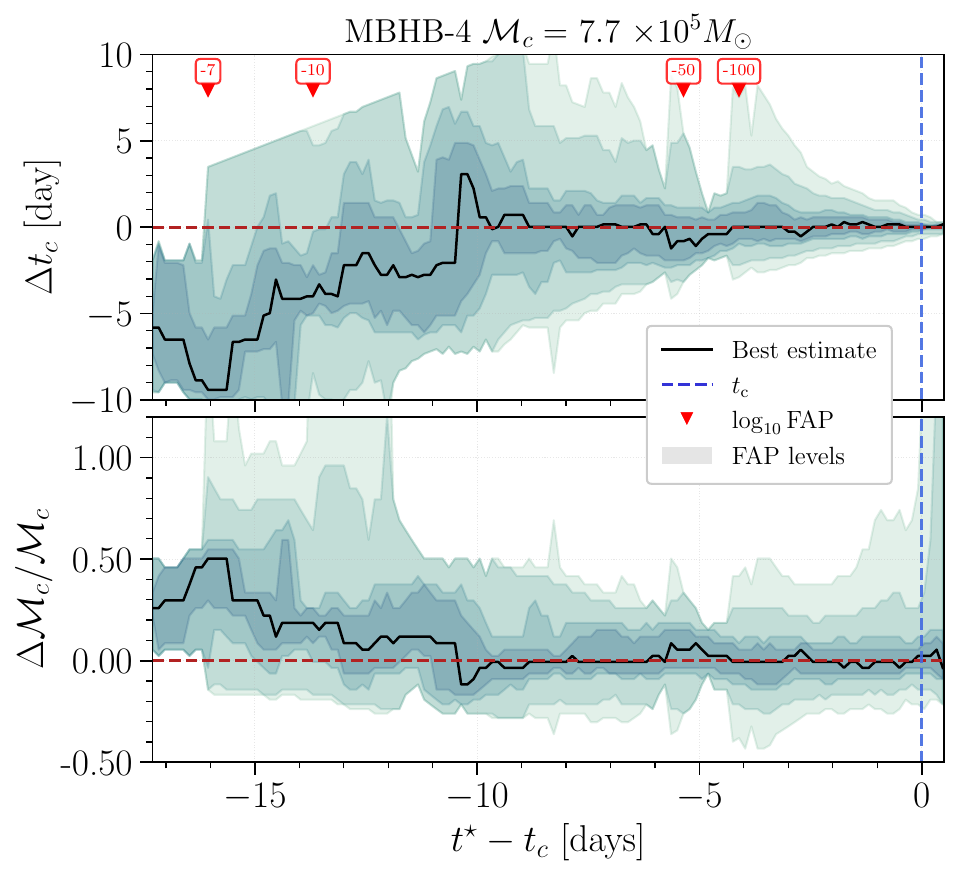}
    \caption{
    Evolution of $\mc$ and $\tc$ estimates for MBHB-4, following conventions in~\cref{fig:FigureA1,fig:FigureA2}. Results are obtained without retaining overlapping points, only. 
    The signal is seen first approximately $\sim 17$ days before merger, whereas enforcing the overlap criterion restricts tracking to about $\sim 8$ days before merger. Nevertheless, during the first $\sim 7$ days of evolution the tracking is affected by biases induced by the overlapping signal MBHB-3.
    }
    \label{fig:FigureA3}
\end{figure*}

The panel corresponding to MBHB-4 shows estimates extending to approximately $\sim 8$ days before merger. In fact, this signal is already detected about $\sim 17$ days before merger; however, a mislabeling issue arises during the tracking of the corresponding region in the FAP map. In~\cref{fig:FigureA3}, we present results obtained with the same analysis without enforcing the retention of only overlapping points in the region-tracking described in~\cref{subsec:coherent_refinement}. 
In this case, the algorithm continuously tracks the signal from the first detection through merger.
Nevertheless, during the first $\sim 7$ days of evolution the tracking is affected by significant biases due to the interaction with the overlapping signal MBHB-3.

Furthermore, we observe a qualitatively different behavior of the FAP contour levels compared to the case in which the retention of only overlapping points is enforced. 
In the present configuration, the activated region is no longer bounded by the initial detection estimates, making the contour levels more susceptible to oscillations. Focusing on the final $\sim 5$ days before merger, we find that all contour levels associated with the estimate of $\tc$ become progressively more constrained with time. 
A different behavior is observed for $\mc$: Although the $90\%$–$95\%$ contours around the FAP minimum remain relatively well confined, the $50\%$–$70\%$ contours broaden toward larger values. This is due to the higher-frequency portion of the signal becomes dominant over the inspiral, close to merger. As a result, the activated regions in the FAP maps become increasingly elongated in the $\mc$ direction while simultaneously tightening in the $\tc$ direction, as illustrated in~\cref{fig:Figure5}. 

Both configurations of the tracking algorithm represent valid alternatives. Enforcing the retention of only overlapping points across successive timesteps yields a more constrained evolution in the $\mc$–$\tc$ space, at the cost of potentially discarding information when the initial estimates are biased. Conversely, allowing regions to be tracked without enforcing overlap produces evolving estimates that more faithfully reflect the instantaneous structure of the FAP maps. While the $\tc$ estimates naturally become more constrained as the merger is approached, a sensible strategy for $\mc$ is to interpret the $90\%$–$95\%$ contours around the FAP minimum as meaningful measures of uncertainty.


\clearpage

\section{Slice-averaged pixel power background distribution}
\label{app:background_distro_derivation}

We provide here a short derivation of results relevant to justify the choice of parametric fit for the background distribution of slice-averaged power, i.e. $\psd \mid {\cal H}_0$.
As detailed in~\cref{sec:methodology}, spectrograms are defined by Hann-windowed short Fourier transforms

\begin{align}
S[p,q] =
\sum_{k=0}^{N-1}
w[k - p h]\, x[k]\, e^{-i 2\pi q k / N}\,, \label{eq:app-sfts}
\end{align}
suitably whitened by the estimator of the quasi-stationary ${\rm PSD}_0[p;t_u]$ in~\cref{eq:psd0qtu}.
For completeness we report here the value of normalization constant for the Hann window: using the trigonometric identity
$\cos^2\theta = \tfrac{1}{2}(1+\cos 2\theta)$
we obtain from~\cref{eq:hannwindow}
\begin{align}
w[n]^2 &=
\frac{1}{4}
\left(
1 - 2\cos\theta_n + \cos^2\theta_n
\right) 
=
\frac{3}{8}
- \frac{1}{2}\cos\theta_n
+ \frac{1}{8}\cos(2\theta_n)\,,\label{eq:app-wofnsquared}\\
\theta_n &= 2\pi n/(N-1)\,,\label{eq:app-thetaofn}
\end{align}
which yields
\begin{align}
\norm{w}^2 &= \sum_{n=0}^{N-1} w[n]^2 = \frac{3}{8}(N-1)\,. \label{eq:app-normwofn}
\end{align}
Consequently, 
$S[p,q]$ is a circular complex Gaussian
random variable with zero mean,
\begin{equation}
\mathbb{E}[S[p,q]] =
\sum_{k=0}^{N-1}
w[k - p h]\,
\mathbb{E}[x[k]]\,
e^{-i 2\pi q k / N}
= 0,
\end{equation}
and second moment

\begin{align}
\mathbb{E}\!\left[ |S[p,q]|^2 \right]
&=
\mathbb{E}\!\left[
\left(
\sum_{k=0}^{N-1}
x[k] w[k - p h] e^{-i 2\pi q k / N}
\right)
\left(
\sum_{\ell=0}^{N-1}
x[\ell] w[\ell - p h] e^{-i 2\pi q \ell / N}
\right)^{\!*}
\right] \\
&=
\sum_{k=0}^{N-1}
\sum_{\ell=0}^{N-1}
w[k - p h] w[\ell - p h]
e^{-i 2\pi q (k-\ell)/N}
\mathbb{E}[x[k]x[\ell]] \\
&=
\sigma_x^2
\sum_{k=0}^{N-1}
w^2[k - p h] = \sigma_x^2 \norm{w}^2\\
&= \frac{3 \sigma_x^2}{8}(N-1) 
\end{align}
where $\mathbb{E}[\cdot]$ denotes the expectation value operator.
The amplitude $R \equiv |S|$ is therefore Rayleigh-distributed, with probability density specified by the parameter $\sigma_r$
\begin{align}
p_R(r; \sigma_r) &=
\frac{r}{\sigma_r^2}
\exp\!\left(-\frac{r^2}{2\sigma_r^2}\right)\,,\label{eq:app-rayleigh1}\\
\sigma_r^2 &= \mathbb{E}\!\left[ |S[p,q]|^2 \right]/2\,.
\end{align}
For reproducibility, we note that~\cref{eq:app-rayleigh1} is consistent with the Rayleigh distribution implemented in \texttt{scipy.stats}, with the scale parameter chosen as
$\sigma_r \norm{w}^2/\sqrt{f_s} = \sigma_x/\sqrt{2f_s}$.

\begin{figure}[t]
    \centering
    \includegraphics[width=\textwidth]{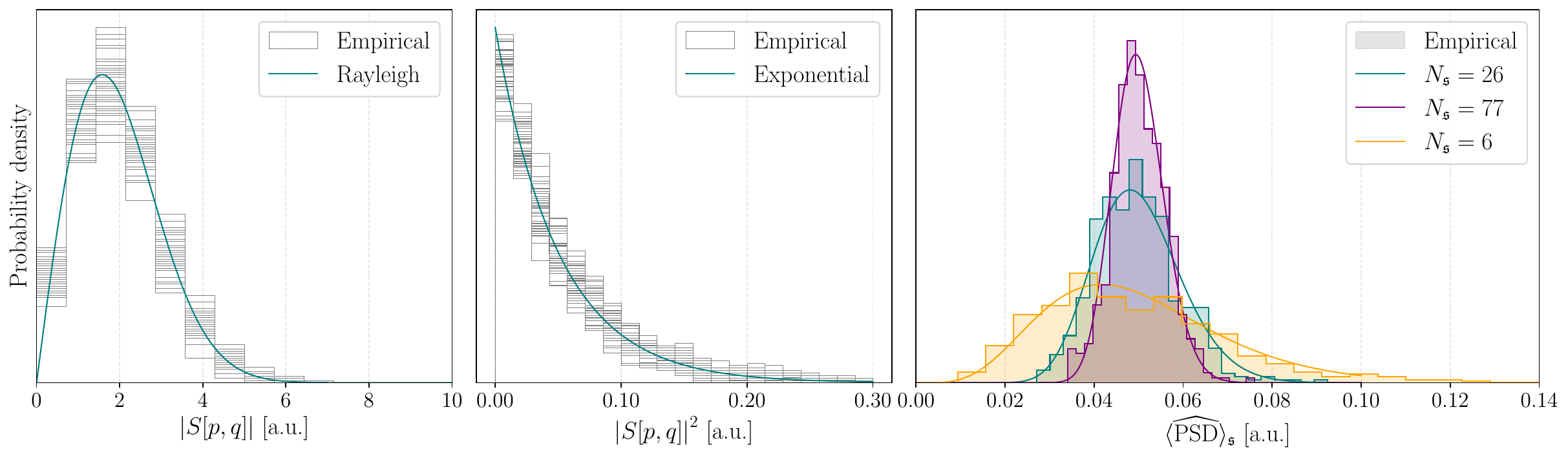}
    \caption{
    Comparisons between analytical models and empirical distributions for  STFT pixel amplitude spectral density (left), power spectral density (middle) and slice averaged power (right).
    The teal curves in left and middle panel denote the Rayleigh and exponential distributions predicted in~\cref{eq:app-rayleigh1,eq:app_exp}, compared against empirical distributions of simulated spectrogram pixels (gray histograms).
    (\textit{Right panel}) Empirical and analytical distributions of slice STFT power averaged over 6 (26,77) pixels in orange (teal, purple), shown as histograms and solid curves, respectively. 
    The prediction of Gamma distributed values in~\cref{eq:app_gamma} justifies our choice of parametric fit shown in~\cref{fig:Figure4}.}    \label{fig:FigureA4}
\end{figure}

\Cref{eq:app-rayleigh1} represents the probability distribution of a single pixel amplitude spectral density, white Gaussian noise realisations. We verify the correctness of our result by generating white Gaussian noise STFTs from non-overlapping segments and collecting samples of a given pixel in each spectrogram. A comparison between the analytical model and the numerically simulated one is shown in~\cref{fig:FigureA4}.
We obtain the distribution of STFT pixel power, including the normalization factor $T_s/\norm{w}^2$ from~\cref{eq:stftpower}, i.e. an exponential distribution with rate parameter $\lambda$:
\begin{align}
p(s; \lambda) &= \lambda e^{-\lambda s} \,\,\, s>0\\\label{eq:app_exp}
\lambda &= \frac{8 \sigma_x^2}{3 N (N-1)} 
\end{align}
or equivalently by the scale parameter $\theta = 1/\lambda$ for the \texttt{scipy.stats} implementation.
To determine how the average slice power in ~\cref{eq:average_asd} is distributed, we define it as 
\begin{align}
p_{\psd}(x) = \int_0^\infty \cdots \int_0^\infty
\delta\!\left(
x - \frac{1}{\Ns}\sum_{i=1}^{\Ns} x_i
\right)
\prod_{i=1}^{\Ns} p(x_i; \lambda)\, {\rm d}x_i, \label{eq:app-laplace}
\end{align}
and obtain its Laplace transform
\begin{align}
\mathcal{L}\{p_{\psd}\}(t)
&=
\int_0^\infty e^{-ts} p_{\psd}(s) \, {\rm d}s \\
&=
\prod_{i=1}^{\Ns}
\int_0^\infty e^{-{t x_i}/{\Ns}} p(x_i;\lambda)\, {\rm d}x_i\\
&=
\left(
\frac{\lambda \Ns}{\lambda \Ns+ t}
\right)^{\Ns}.
\end{align}
Through inverse Laplace transform we recover the explicit expression for the probability distribution, 
\begin{align}
p_{\psd}(s; \lambda, \Ns) = \frac{(\lambda \Ns)^{\Ns}
   }{\Gamma
   (\Ns)}s^{\Ns -1} e^{-s \lambda \Ns},
\qquad s \ge 0\,, \label{eq:app_gamma}
\end{align}
which is a Gamma distribution with shape parameter $\Ns$ and scale parameter $1/{\lambda\Ns}$. This result supports the choice of a Gamma distribution as a parametric fit in~\cref{subsec:detection_strategy}.
Nonetheless, we point out that the chosen STFT therein uses overlapping window segments with fixed hop size $h<N$, thereby violating the independence of pixels assumed in~\cref{eq:app-laplace}, while preserving individual ones Gaussianity.
We account for it computing the covariance between two pixels specified by $p,q$ and $p^\prime,q^\prime$, respectively, i.e.

\begin{align}
\mathbb{E}\!\left[ S[p,q] S[p',q']^\ast \right] &= \mathbb{E}\!\Bigg[
\left(
\sum_{k=0}^{N-1}
w[k - p h] x[k] e^{-i 2\pi q k / N}
\right)
\left(
\sum_{\ell=0}^{N-1}
w[\ell - p' h] x[\ell] e^{-i 2\pi q' \ell / N}
\right)^{\!*}
\Bigg] \\
&= \sum_{k=0}^{N-1}
\sum_{\ell=0}^{N-1}
w[k - p h] w[\ell - p' h]
e^{-i 2\pi (q k-q^\prime \ell) / N}
\mathbb{E}[x[k] x[\ell]^\ast]\\
&= \sigma_x^2
\sum_{k=0}^{N-1}
w[k - p h] w[k - p' h]
e^{-i 2\pi (q - q') k / N}\,.\label{eq:app-correlation}
\end{align}

A complete expression for the covariance can be obtained numerically for specific windows and segment overlaps. 
To take such correlation into account, we note that our aim is to construct the survival function for the slice-averaged pixel power.
As just shown, pixel STFT values of a slice $\mathfrak{s}$ will be jointly-distributed according to a multivariate complex Gaussian, with covariance $\Sigma_{\mathfrak s}$, i.e.
\begin{equation}
 \mathbf{s} = \lbrace S[p_i,q_i], (p_i,q_i) \in \mathfrak{s} \rbrace \sim \mathcal{CN}(0, \Sigma_\mathfrak{s})\,.
\end{equation}
Once a slice is chosen, the covariance can be diagonalised through eigenvalue decomposition $\Sigma_{\mathfrak{s}} = U \Lambda U^\dagger$, where $U$ is unitary and $\Lambda = \mathrm{diag}(\lambda_1, \ldots,\lambda_{\Ns})$ is the matrix of eigenvalues.
The detection statistic $\psd$ in~\cref{eq:average_asd} is defined as the squared norm of $\norm{\mathbf{s}}^2 = \bm{s}^\dagger \bm{s}$, scaled by a conventional normalization constant. 
Knowing that the auxiliary variables $\bm{y} = U^\dagger \bm{s}$
are uncorrelated complex Gaussian random
variables,
\begin{equation}
y_i \sim \mathcal{CN}(0, \sigma_i^2),
\qquad
\sigma_i^2 = \lambda_i\,,
\end{equation}
and that unitary transformations preserve the squared norm
\begin{align}
\psd \propto \norm{\bm{s}}^2 &= \norm{\bm{y}}^2 = \sum_{i=1}^{\Ns} y_i^2\,,
\end{align}
we finally obtain for the survival function 
\begin{align}
{\rm Prob}\left(\psd > s \mid {\cal H}_0\right)
=
\sum_{k=1}^{\Ns}
\left(
\exp\!\left(-\frac{s}{\sigma_k^2}\right)
\prod_{\substack{j=1 \\ j \neq k}}^{\Ns}
\frac{\sigma_j^2}{\sigma_j^2 - \sigma_k^2}
\right),
\qquad s \ge 0.
\end{align}
\twocolumngrid
\end{document}